\documentclass[aps,prx,10pt,twocolumn,superscriptaddress,nofootinbib]{revtex4-2}
\pdfoutput=1

\usepackage[ascii]{inputenc}
\usepackage[english]{babel}
\usepackage[T1]{fontenc}
\usepackage{hyperref}

\usepackage{graphicx}
\usepackage[dvipsnames]{xcolor}
\usepackage{scrhack} 
\usepackage{listings}
\usepackage{lstautogobble}
\usepackage{enumitem}

\usepackage[textsize=tiny]{todonotes}

\newcommand{\tw}{\text{tw}}

\usepackage{algorithmic}
\usepackage[ruled]{algorithm2e}
\usepackage{amsmath}

\usepackage{subcaption}
\usepackage{lipsum}

\usepackage{amsfonts}

\usepackage{amssymb}
\usepackage{amsthm}
\newtheorem{definition}{Definition}

\newtheorem{theorem}{Theorem}
\usepackage{float}
\usepackage{cases}

\newcommand{\cg}[2]{\begingroup\color{#1}#2\endgroup}

\newcommand{\C}[0]{\mathcal{C}}

\definecolor{darkblue}{rgb}{0.0, 0.0, 0.55}
\definecolor{trueblue}{rgb}{0.0, 0.45, 0.81}
\definecolor{persimmon}{rgb}{0.93, 0.35, 0.0}

\newcommand{\cf}[2]{\frac{\begingroup\color{darkblue}#1\endgroup}{\begingroup\color{persimmon}#2\endgroup}}

\lstdefinestyle{base}{
  language=C,
  emptylines=1,
  breaklines=true,
  moredelim=**[is][\color{blue}]{@}{@},
}

\makeatletter 
    
\renewcommand\onecolumngrid{
\do@columngrid{one}{\@ne}%
\def\set@footnotewidth{\onecolumngrid}
\def\footnoterule{\kern-6pt\hrule width 1.5in\kern6pt}%
}

\renewcommand\twocolumngrid{
        \def\footnoterule{
        \dimen@\skip\footins\divide\dimen@\thr@@
        \kern-\dimen@\hrule width.5in\kern\dimen@}
        \do@columngrid{mlt}{\tw@}
}%

\makeatother

\begin{document}

\title{On the Optimal Linear Contraction Order of Tree Tensor Networks, and Beyond}

\author{Mihail Stoian}
\email{mihail.stoian@tum.de}
\affiliation{Technische Universit\"at M\"unchen, School of CIT, Department of Computer Science, Boltzmannstra{\ss}e 3, 85748 Garching, Germany}
\author{Richard M.~Milbradt}
\email{r.milbradt@tum.de}
\affiliation{Technische Universit\"at M\"unchen, School of CIT, Department of Computer Science, Boltzmannstra{\ss}e 3, 85748 Garching, Germany}
\author{Christian B.~Mendl}
\email{christian.mendl@tum.de}
\affiliation{Technische Universit\"at M\"unchen, School of CIT, Department of Computer Science, Boltzmannstra{\ss}e 3, 85748 Garching, Germany}
\affiliation{Technische Universit\"at M\"unchen, Institute for Advanced Study, Lichtenbergstra{\ss}e 2a, 85748 Garching, Germany}

\begin{abstract}
The contraction cost of a tensor network depends on the contraction order. However, the optimal contraction ordering problem is known to be NP-hard.~We show that the \emph{linear} contraction ordering problem for \emph{tree} tensor networks admits a polynomial-time algorithm, by drawing connections to database join ordering.~The result relies on the adjacent sequence interchange property of the contraction cost,~which enables a global decision of the contraction order based on local comparisons.~Based on that, we specify a modified version of the IKKBZ database join ordering algorithm to find the optimal tree tensor network linear contraction order. Finally, we extend our algorithm as a heuristic to general contraction orders and arbitrary tensor network topologies.
\end{abstract}

\maketitle

\let\oldaddcontentsline\addcontentsline 
\renewcommand{\addcontentsline}[3]{} 


\section{Introduction}\label{sec:introduction}

Originally developed for use in theoretical physics~\cite{Schollwoeck2011} and in parallel in mathematics~\cite{hackbusch}, tensor networks are nowadays an active interdisciplinary field of research, which led to fruitful interchanges and insights between disciplines~\cite{MAL-059, MAL-067, quantum_via_tn, Stoudenmire2016SupervisedLW, Szalay2015, khamis_faq, dudek2020efficient}. From their inception, questions of computational complexity and the feasibility of efficient manipulation of tensor networks played a major role.

In this work, we are concerned with an efficient network contraction algorithm, which can be thought of as a stepwise transformation of a tensor network into a single tensor. A single step consists of contracting two tensors. This step is repeated until only one tensor remains. Although the final result does not depend on the order in which the pairwise tensor contractions are performed, the overall performance is significantly affected. For this reason, one aims to find an optimal contraction order before implementing the actual contraction.

However, finding the optimal order for a generic network of $n$ tensors is an NP-hard problem~\cite{Lam1997OnOA}. Therefore, one must settle for the exponential algorithm for small instances, e.g., $n \leq 20$, and hope for good contraction orders via heuristics otherwise, or focus on tensor network classes for which the problem becomes tractable. In this paper, we specify a polynomial-time algorithm for finding the optimal \emph{linear} contraction order of a \emph{tree} tensor network.

\textbf{\normalfont\bfseries Prior Work.} Pfeifer et al.~\cite{dp_in_tensor_network} have devised a graph-theoretic dynamic program for finding optimal contraction sequences, which inherently takes exponential time. More recently, Gray and Kourtis~\cite{cotengra} have introduced a novel optimizer based on hypergraph partitioning that has been shown to yield high-quality contraction orders. Another line of research has focused on parallel contraction algorithms for tensor networks representing quantum circuits, introducing two ad-hoc methods that refine the final contraction order, namely dynamic slicing~\cite{chen2018classical} and local optimization~\cite{huang2020classical}. However, to find an \emph{initial} order, they still rely on heuristics that do not provide an optimality guarantee.

The closest work to ours is the optimal polynomial-time algorithm of Xu et al.~\cite{ttn_optimal_max}, which minimizes the \emph{time} complexity of a tree tensor network contraction in terms of big-$\mathcal{O}$ notation. Note that their cost function is only a proxy for the actual cost function $\C$, defined in Eq.~\eqref{eq:contr_cost_rec} below, which sums up \emph{all} intermediate contraction costs. Our optimal algorithm \texttt{TensorIKKBZ} directly targets $\C$. Recently, Ibrahim et al.~\cite{cubic_dp_in_tensor} have optimized the contraction tree shape of a generic tensor network given a \emph{fixed} permutation of the tensors via the well-known dynamic program from the matrix-chain multiplication problem~\cite{cormen}. However, they rely on heuristics to find the initial permutation. In contrast, adapting a novel database join ordering technique~\cite{adaptive}, we propose to use \texttt{TensorIKKBZ}'s \emph{optimal} linear contraction order as a seed for the initial permutation.

We are not the first to point out the similarity between tensor network contraction ordering and database join ordering. The first work we know of is by Khamis et al.~\cite{khamis_faq}, which combines several problems from database theory, tensor networks, and probabilistic graphical models under the same umbrella. Later, Dudek et al.~\cite{dudek2020efficient} reiterated this similarity and provided contraction orders of small maximum rank via tree decompositions. Indeed, tree decompositions are a common tool for optimizing contraction orders: Markov and Shi~\cite{shi_tree_decomp} use them to simulate quantum circuits of $T$ gates, with underlying circuit graph $G$, in time $T^{\mathcal{O}(1)} \exp(\mathcal{O}(\tw(G)))$, where $\tw(G)$ is the tree-width of $G$. (Tree-width is a measure of how tree-like a graph is~\cite{Cygan_book}.) However, the cost function they use is slightly different: They define the \emph{contraction complexity} of a contraction order as the \emph{maximum} degree of a merged tensor during the contraction. This, however, optimizes only for the \emph{size} of the largest possible intermediate tensor, ignoring the cumulative contribution of intermediate tensor sizes, which may be larger by a multiplicative factor of $n - 1$.\footnote{There are $n - 1$ pair-wise contraction steps in total.} This is the case for all other prior research that relies on tree decompositions~\cite{boixo2018simulation, roman_tree_decomp}.

\textbf{\normalfont\bfseries Contribution.} Our main technical result is an optimal algorithm for \emph{linear} contraction orders of \emph{tree} tensor networks (Sec.~\ref{sec:algorithm}). We further build upon our optimal algorithm and consider general contraction orders and generic tensor networks as well, for which we provide near-optimal contraction orders (Sec.~\ref{sec:beyond}).


\section{Background}\label{sec:background}

\begin{figure}[H]
    \centering
    \includegraphics{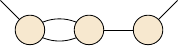}
    \caption[Example Tensor Network]{A network with three tensors}
    \label{fig:example_tn}
\end{figure}

In this work, a tensor is an element of $\mathbb{C}^{n_1 \times \ldots \times n_d}$, where $d$ is called the degree. A tensor network is defined as a set of $n$ tensors, where the tensors are represented as vertices and the legs along which the tensors are to be contracted are represented as edges. Note that some tensors may have \emph{open} (uncontracted) legs, corresponding to edges with only one endpoint. In Fig.~\ref{fig:example_tn}, we illustrate a tensor network with three tensors.

\subsection{Tensor Contraction}

The contraction of $T^{[1]}\in\mathbb{C}^{p_1\times\ldots\times p_a \times q_1\times \ldots\times q_b}$ with $T^{[2]}\in \mathbb{C}^{q_1\times\ldots\times q_b \times r_1\times\ldots\times r_c}$ gives another tensor $T^{[1,2]}\in\mathbb{C}^{p_1\times\ldots\times p_a \times r_a\times\ldots\times r_c}$, defined as \begin{equation}
T^{[1,2]}_{i_1,\ldots,i_a,k_1,\ldots,k_c}=\displaystyle\sum_{j_1,\ldots,j_b} T^{[1]}_{i_1,\ldots,i_a,j_1,\ldots,j_b} T^{[2]}_{j_1,\ldots,j_b,k_1,\ldots,k_c}.
\label{eq:tensor_contraction} 
\end{equation}
The tensor contraction cost is then defined as
\begin{equation}
c\left(T^{[1]}, T^{[2]}\right) = \prod_{i=1}^{a} p_i\prod_{j=1}^{b} q_j\prod_{k=1}^{c} r_k,
\label{eq:tensor_contraction_cost}
\end{equation}
i.e., the product of the sizes of all legs involved in the operation, which is exactly equal to the number of scalar multiplications required for the contraction.\footnote{This cost, already established in the literature, assumes a na{\"i}ve implementation of the operation, i.e., nested loops over the dimensions. Especially for matrix-matrix multiplication, Strassen's algorithm and its refinements achieve a lower computational complexity; we do not consider these here for simplicity.}


\subsection{Contraction Order}\label{subsec:co}

\begin{figure*}
    \centering
    \begin{subfigure}[c]{0.3\columnwidth}
        \centering
        \includegraphics[]{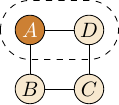}
        \phantomsubcaption
    \end{subfigure}$\rightarrow$
    \begin{subfigure}[c]{0.3\columnwidth}
        \centering
        \includegraphics[]{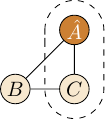}
        \phantomsubcaption
    \end{subfigure}$\rightarrow$
    \begin{subfigure}[c]{0.285\columnwidth}
        \centering
        \includegraphics[]{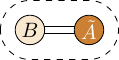}
        \phantomsubcaption
    \end{subfigure}$\rightarrow$
    \begin{subfigure}[c]{0.115\columnwidth}
        \centering
        \includegraphics[]{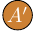}
        \phantomsubcaption
    \end{subfigure}
\caption[Linear Contraction Order]{Linear contraction order of a tensor network with four tensors}
\label{fig:example_linear_contraction}
\end{figure*}

\begin{figure*}
    \centering
    \begin{subfigure}[c]{0.3\columnwidth}
        \centering
        \includegraphics[]{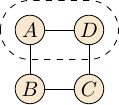}
        \phantomsubcaption
    \end{subfigure}$\rightarrow$
    \begin{subfigure}[c]{0.3\columnwidth}
        \centering
        \includegraphics[]{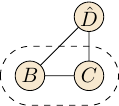}
        \phantomsubcaption
    \end{subfigure}$\rightarrow$
    \begin{subfigure}[c]{0.225\columnwidth}
        \centering
        \includegraphics[]{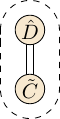}
        \phantomsubcaption
    \end{subfigure}$\rightarrow$
    \begin{subfigure}[c]{0.175\columnwidth}
        \centering
        \includegraphics[]{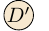}
        \phantomsubcaption
    \end{subfigure}
\caption[General Contraction Order]{General contraction order of a tensor network with four tensors}
\label{fig:example_general_contraction}
\end{figure*}

While the result of a tensor network contraction is independent of  the order in which the individual tensors are contracted, the execution time is highly dependent on the chosen order. There are two types of contraction orders, \emph{linear} and \emph{general}, as classified by O'Gorman~\cite{linear_and_general}. A linear contraction order can only contract a fixed tensor iteratively with the others, while a general one is allowed to contract the tensors in an arbitrary order. To exemplify their difference, let us consider the network contraction in Fig.~\ref{fig:example_linear_contraction}. There, we contract the tensor $A$ in each step with one of its neighbors. Thus, this is a \emph{linear} contraction order. In contrast, Fig.~\ref{fig:example_general_contraction} shows an example of a \emph{general} contraction order for the same tensor network.

Due to their flexibility, general orders can provide lower contraction costs as compared to linear orders. However, due to the exponentially larger search space, they are also more difficult to optimize.


\subsection{Contraction Tree}\label{subsec:contraction_tree}

Complementary to contraction orders, \emph{contraction trees}~\cite{contraction_tree_first} offer an intuitive visualization of the order in which the contractions are performed. Formally, a contraction tree is a rooted binary tree. Its leaves are the tensors of the network, and the internal nodes correspond to the contractions between the tensors represented by the left and right children, respectively.

\begin{figure}
    \centering
    \begin{subfigure}[b]{0.4\columnwidth}
        \centering
        \includegraphics[]{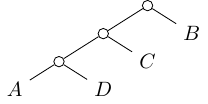}
        \caption{Linear}
    \end{subfigure}
    \begin{subfigure}[b]{0.4\columnwidth}
        \centering
        \includegraphics[]{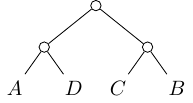}
        \caption{General}
    \end{subfigure}
    \caption[Contraction Tree Types]{Contraction tree types}
    \label{fig:contraction_trees}
\end{figure}

We depict in Fig.~\ref{fig:contraction_trees} the contraction trees associated with the contraction orders in Fig.~\ref{fig:example_linear_contraction} and Fig.~\ref{fig:example_general_contraction}, respectively. For convenience, we refer to a contraction tree associated with a linear/general contraction order as a linear/general contraction tree.

\subsection{Contraction Cost}\label{subsec:contraction_cost}

The contraction cost $\C$ of the entire network is the sum of all intermediate contraction costs. This can be recursively defined on the contraction tree as
\begin{equation}
\mathcal{C}(T) =
\begin{cases}
    0, & \text{if } T \text{ is a single tensor} \\
    \mathcal{C}(T^{[a]}) + \mathcal{C}(T^{[b]}) & \\
    \,\, +\:c(T^{[a]}, T^{[b]}), & \text{if } T = T^{[a]}T^{[b]}.
\end{cases}\label{eq:contr_cost_rec}
\end{equation}
Alternatively, it can be explicitly defined as the sum of the contraction costs at each inner node of the contraction tree~\cite{cotengra}. We prefer the recursive definition because it is easier to handle in the proofs. An example of its calculation for the contraction order shown in Fig.~\ref{fig:example_linear_contraction} is given in Appendix~\ref{appendix:contraction_cost}.


\section{Algorithm}\label{sec:algorithm}

First introduced by Monma and Sidney~\cite{Monma1979SequencingWS} for the problem of scheduling with precedence constraints, and then adapted to database join ordering by Ibaraki and Kameda~\cite{ik} and improved by Krishnamurthy et al.~\cite{kbz}, the \texttt{IKKBZ} algorithm\footnote{Its name is derived from the initials of its authors.} operates only on \emph{tree} queries and returns the optimal \emph{linear} join tree without relational cross products, assuming that the cost function has the \emph{adjacent sequence interchange} (ASI) property~\cite{Monma1979SequencingWS}.

The language of database query optimization can be translated into the context of tensor networks (see Sec.~\ref{sec:query_optimization} for details). In the following, we show that the cost function used in tensor networks, Eq.~\eqref{eq:contr_cost_rec}, does indeed satisfy the ASI property, which allows us to use \texttt{IKKBZ} for tensor network contraction ordering.


\subsection{Precedence Graph}\label{subsec:precedence_graph}

\begin{figure}
    \centering
    \includegraphics{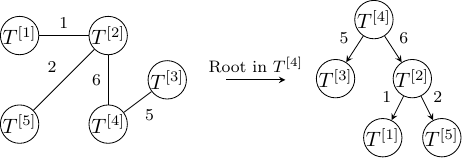}
    \caption[Precedence Graph]{Precedence graph of $T^{[4]}$ (right) is obtained by rooting the tree tensor network (left) in $T^{[4]}$}
    \label{fig:pg}
\end{figure}

The main idea of the original algorithm comes from the observation that we are working only with linear orders. In a linear order, we start with a tensor and then sequentially contract it with neighboring tensors (cf. Sec.~\ref{subsec:co}). Moreover, since we are working with tree-shaped networks, this naturally imposes a \emph{precedence} relation on the tensors of the network. Therefore, we can consider each tensor as the first element of the solution and thus root the tree in it. The subsequent tensors to be contracted must obey the precedence relation, i.e., a child node must not appear in the contraction order before its parent. An example of a precedence graph of a tree tensor network is depicted in Fig.~\ref{fig:pg}. In that example, the precedence graph enforces that $T^{[4]} \rightarrow T^{[3]}$ and $T^{[4]} \rightarrow T^{[2]}$, i.e., $T^{[4]}$ should appear in the contraction order before $T^{[3]}$ and $T^{[2]}$, respectively; similarly for $T^{[2]} \rightarrow T^{[1]}$ and $T^{[2]} \rightarrow T^{[5]}$.

To achieve optimality, the algorithm roots the tree in each node, solves the ordering problem for the obtained rooted tree, and chooses the order that leads to the minimum cost. 

\subsection{Adjacent Sequence Interchange Property}\label{sec:asi_property}

For its optimality, the algorithm requires the cost function to satisfy the ASI property, defined in the following:

\begin{definition}[ASI Property~\cite{Monma1979SequencingWS}]\label{def:asi}
Let $A$ and $B$ be two sequences and $U$ and $V$ two non-empty sequences. We say that a cost function $C$ has the \emph{adjacent sequence interchange (ASI)} property, if and only if there exists a score function $\sigma$ such that the \mbox{following} holds:
\[
C(AUVB) \leq C(AVUB) \iff \sigma(U) \leq \sigma(V),
\]
if $AUVB$ and $AVUB$ satisfy the precedence constraints imposed by a given precedence graph.\footnote{The join ordering literature uses the term \emph{rank} instead of \emph{score}. However, to avoid any confusion with the rank of a matrix, we decided to use the term \emph{score} (in the same mathematical sense).}
\end{definition}

Thus, a cost function $C$ that satisfies the ASI property allows us to use \emph{local} comparisons based on the score function, i.e., $\sigma(U) \leq \sigma(V)$, to infer comparisons between \emph{global} costs, i.e., $C(AUVB) \leq C(AVUB)$. The \texttt{IKKBZ} algorithm makes extensive use of score comparisons to find the optimal order.


\subsubsection{ASI Exemplification}\label{sec:asi_motivation}

Consider the tree tensor network and one of its precedence graphs (that of $T^{[1]}$) in Fig.~\ref{fig:asi_example_pqr}. 
Let us (manually) determine which of the following linear contraction orders is better by computing the costs using Eq.~\eqref{eq:contr_cost_rec} (a complete derivation can be found in Appendix~\ref{appendix:asi_derivation}):
\begin{figure}
    \centering
    \includegraphics{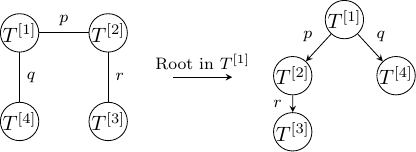}
    \caption{Precedence graph of $T^{[1]}$, where $T^{[1]}\in\mathbb{C}^{p \times q}, T^{[2]}\in\mathbb{C}^{p \times r}, T^{[3]}\in\mathbb{C}^{r}$, and $T^{[4]}\in\mathbb{C}^{q}$}
    \label{fig:asi_example_pqr}
\end{figure}
\begin{flalign*}
\mathcal{C}(T^{[1]}\cg{trueblue}{T^{[2]}}\cg{persimmon}{T^{[4]}}T^{[3]}) &= pqr + qr + r\\
\mathcal{C}(T^{[1]}\cg{persimmon}{T^{[4]}}\cg{trueblue}{T^{[2]}}T^{[3]}) &= pq + pr + r.
\end{flalign*}
When comparing both costs, i.e.,
\begin{flalign*}
    &\qquad& pqr + qr + r &\leq pq + pr + r &&\\
    \iff&& pqr + qr &\leq pq + pr &&\\
    \iff&& pr(q - 1) &\leq q(p - r),
\end{flalign*}
\label{eq:naive} we observe that the original inequality reduces to a simpler one that only considers the tensors $\cg{trueblue}{T^{[2]}}$ and $\cg{persimmon}{T^{[4]}}$. To see this, assume $q > 1$ and $p > r$, such that the inequality sign is maintained. Then the inequality reduces to $\frac{pr}{p - r} \leq \frac{q}{q - 1}$. Indeed, $p$ and $r$ are the sizes of the legs of $\cg{trueblue}{T^{[2]}}$, while $q$ is the size of the only leg of $\cg{persimmon}{T^{[4]}}$. Hence, a \emph{local} comparison is enough to decide whether $T^{[1]}\cg{trueblue}{T^{[2]}} \cg{persimmon}{T^{[4]}} T^{[3]}$ is better than $T^{[1]}\cg{persimmon}{T^{[4]}} \cg{trueblue}{T^{[2]}} T^{[3]}$.


\subsubsection{Cost Function Reformulation}\label{subsec:simplifications}

Before proving that $\mathcal{C}$ has the ASI property, let us simplify the definition of $\mathcal{C}$, since we are dealing only with \emph{linear} orders and \emph{tree} tensor networks. First, we reformulate the tensor contraction cost given in Eq.~\eqref{eq:tensor_contraction_cost}. To this end, we assign to each tensor $T\in\mathbb{C}^{n_1\times\ldots\times n_d}$ a \emph{size}, which is the product of its leg sizes, i.e., $\#T = \prod_{i=1}^{d} n_i$. We have
\begin{equation}
\begin{split}
c\left(T^{[1]}, T^{[2]}\right) &= \prod_{i=1}^{a} p_i\prod_{j=1}^{b} q_j\prod_{k=1}^{c} r_k\\
&= \frac{\prod_{i=1}^{a} p_i\prod_{j=1}^{b} q_j\prod_{j=1}^{b} q_j\prod_{k=1}^{c}r_k}{\prod_{j=1}^{b}q_j}\\
&= \frac{\#T^{[1]}\#T^{[2]}}{\prod_{j=1}^{b}q_j}.
\label{eqn:reformulation}
\end{split}
\end{equation}
This reformulation, already presented by Evenbly and Pfeiffer~\cite[Eq.~(4)]{contraction_tree_first}, is the key ingredient for the connection to database join ordering, as we will see later in Sec.~\ref{subsec:query_graph}.

The ASI property (Def.~\ref{def:asi}) operates on sequences, which specify the order in which tensors are contracted via a permutation $\pi$, and satisfy the constraints imposed by the precedence graph.\footnote{Sequences and linear orders are the same object.} In particular, if a sequence contains all tensors, its first tensor, $T^{[\pi_1]}$, is the root of the precedence graph. Moreover, since we deal with tree tensor networks and enforce that no outer products emerge, each contraction takes place over a \emph{single} leg.

Intuitively, we can recursively compute the contraction cost of $S$ using a split $S = S_1S_2$ of the sequence into two subsequences. In particular, the cost $\C(S)$ is independent of the choice of the split, as the order of the tensors in the sequence remains the same (we illustrate this in Appendix~\ref{appendix:recursive_example}). This lets us rewrite the definition of $\C$ for sequences as follows, where $\varnothing$ denotes the empty sequence:
\begin{equation}
\C(S) = 
\begin{cases}
    0, & S = \varnothing\\
    0, & S = T^{[\pi_1]}\\
    \#T^{[\pi_i]}, & S = T^{[\pi_i]},\:i \neq 1\\
    \C(S_1) + \frac{\overline{\#}T^{[S_1]}}{|e_{S_1,S_2}|}\C(S_2), & S = S_1 S_2,
\label{eq:other_cout}
\end{cases}
\end{equation}
where $T^{[S]}$ is the tensor resulting from the contraction of the tensors in $S$, $\overline{\#}{T^{[S]}}$ is the size of $T^{[S]}$ \emph{without} the (single) leg to its parent in the precedence graph, and $|e_{S_1,S_2}|$ is the size of the (single) leg between $T^{[S_1]}$ and $T^{[S_2]}$. Note that in a precedence graph this directed edge will always lead to the first tensor occurring in $S_2$. Thus, when it is clear from the context, we will write $e_{S_2}$ instead of $e_{S_1,S_2}$. With this notation, we can rewrite $\overline{\#}{T^{[S]}}$ as $\frac{\#T^{[S]}}{|e_{S}|}$, hence we refer to $\overline{\#}$ as the \emph{capped size}. We exemplify Eq.~\eqref{eq:other_cout} in Appendix~\ref{appendix:recursive_example}.

Before stating the main result, let $\mathbb{S}$ be the set of non-empty sequences observing the constraints imposed by the precedence graph, from which we exclude the sequences starting with the root of the precedence graph. This exception is due to the fact that the algorithm itself does not need to compute the scores of these sequences,\footnote{This is reflected in the if-condition on line 9 in Alg.~\ref{algo:tensor_ikkbz}.} thus ensuring that all $\C$-values will be positive:

\begin{theorem}\label{theorem:asi_cost_function} $\mathcal{C}$ satisfies the ASI property for the score function $\sigma \colon \mathbb{S} \to \mathbb{N}^+ \times \mathbb{Z}$ defined as
\[
\sigma(S) = (\C(S), |e_{S}| - \overline{\#}{T^{[S]}}).
\] In particular, we define ``$\leq$'' in Def.~\ref{def:asi} as
\[
    \sigma(U) \leq \sigma(V) \iff \sigma_1(U) \sigma_2(V) \leq \sigma_2(U) \sigma_1(V),
\] where we refer to $\sigma_1(S)$ and $\sigma_2(S)$ as the numerator and denominator, respectively, of the symbolic fraction $\sigma(S)$.
\end{theorem}
Appendix~\ref{appendix:proof_theorem_1} contains a proof of the theorem. Additionally, in Appendix~\ref{appendix:proof_transitivity} we prove the transitivity of ``$\leq$'', which is implicitly used by the algorithm to ensure correct sorting.

\begin{algorithm*}
  \caption{\texttt{TensorIKKBZ}}\label{algo:tensor_ikkbz}
  \begin{algorithmic}[1]
  \STATE \textbf{Input:} Tree tensor network $\mathcal{T} = (V, E, c)$
  \STATE \textbf{Output:} Optimal linear contraction order $S^*$
    
  \STATE $S^* \leftarrow \boldsymbol\varnothing$
  \FOR{\textbf{each} $T^{[i]}$ in $V$}
      \STATE Build the precedence graph $\mathcal{P}_{T^{[i]}}$ (cf. Sec.~\ref{subsec:precedence_graph})
      \WHILE {$\mathcal{P}_{T^{[i]}}$ is not linearized}
            \STATE Select $T^{[u]}$ whose children are linearized
            \STATE Interleave the linearizations of $T^{[u]}$'s children (by ascending $\sigma$-value)
            \IF{$T^{[u]}\neq T^{[i]}$}
                \STATE Call \textsc{Normalization}\:(Alg.~\ref{algo:normalization}) on the obtained linearization of $T^{[u]}$    
            \ENDIF   
      \ENDWHILE
      \IF{$S^* = \varnothing$ or $\C(\mathcal{P}_{T^{[i]}}) < \C(S^*)$}
                \STATE $S^* \leftarrow \mathcal{P}_{T^{[i]}}$ 
        \ENDIF
  \ENDFOR
  \RETURN\!$S^*$
  \end{algorithmic}
\end{algorithm*}
\begin{algorithm}
  \caption{\texttt{TensorIKKBZ}: \textsc{Normalization}}\label{algo:normalization}
  \begin{algorithmic}[1]
    \STATE \textbf{Input:} Tensor $T$, subtree linearization $\mathcal{L}$
    \STATE \textbf{Output:} Normalized linearization $\mathcal{L}$
    \STATE $i \leftarrow 1$
    \WHILE{$i \leq |\mathcal{L}| 
 \land \sigma(T) \geq \sigma(\mathcal{L}_i)$}
        \STATE Update the numerator of $\sigma(T)$:\\$\sigma_1(T) \leftarrow \sigma_1(T) + \frac{\overline{\#}{T}}{|e_{\mathcal{L}_i}|}\sigma_1(\mathcal{L}_i)$ (cf. Eq.~\eqref{eq:other_cout})
        \STATE Update the denominator of $\sigma(T)$:\\$\sigma_2(T) \leftarrow |e_T| - \frac{\overline{\#}{T_{\mathcal{L}_i}}}{|e_{\mathcal{L}_i}|}\overline{\#}{T}$ (cf. Eq.~\eqref{eq:norm_identity})
        \STATE Fuse $T$ with $\mathcal{L}_i$ as a \emph{compound} tensor
        \STATE $i \leftarrow i + 1$
    \ENDWHILE
  \end{algorithmic}
\end{algorithm}


\subsection{\normalsize\texttt{TensorIKKBZ}}\label{sec:tensor_ikkbz}
We adapt the \texttt{IKKBZ} algorithm to tensor networks, which we refer to as \texttt{TensorIKKBZ}. Intuitively, the algorithm maintains the invariant that the tensors are sorted based on the score function, $\sigma$.\footnote{Ties are arbitrarily broken, preserving precedence constraints.} However, the precedence constraints must also be observed. This can lead to a \emph{contradictory} sequence that violates the invariant: the precedence graph requires $A \rightarrow B$, but $\sigma(A) \geq \sigma(B)$. To restore the invariant, the algorithm merges $A$ and $B$ into a single node. This node is called a \emph{compound tensor} and comprises all the tensors in $A$ and $B$ (which may themselves be compound tensors) in that order. This step is called \emph{normalization}.

To see why normalization does not preclude optimality, consider the following theorem (see~\cite[Thm. 2]{Monma1979SequencingWS}):

\begin{theorem}\label{theorem:normalize}
Let $\{A,B\}$ be a compound tensor. If $A \rightarrow B$ and $\sigma(A) \geq \sigma(B)$, then there exists an optimal sequence in which $B$ directly follows $A$.
\end{theorem}

Appendix~\ref{appendix:proof_theorem_2} contains a proof of this theorem.


\subsubsection{Pseudocode}

We outline the pseudocode of the \texttt{TensorIKKBZ} algorithm in Alg.~\ref{algo:tensor_ikkbz}, which takes the tree tensor network $\mathcal{T}$ as input and outputs the optimal linear contraction order $S^*$. For each tensor $T^{[i]}$ it builds the precedence graph $\mathcal{P}_{T^{[i]}}$ (line 5) and linearizes it (lines 6--12). This is done by repeatedly interleaving subtree linearizations by ascending $\sigma$-value and resolving contradictory sequences therein. Once the entire precedence graph has been linearized, its cost is computed and compared to the best one obtained so far (line~13). Finally, the optimal order is returned (the compound tensors are resolved into their constituent tensors in a step known as denormalization).

Before outlining Alg.~\ref{algo:normalization}, let us consider an identity for the capped size which will be useful in the algorithm when computing scores of compound tensors. Let $S_1$ and $S_2$ be two adjacent subsequences. Then, the capped size of $T^{[S_1,S_2]}$, i.e., the tensor resulting from the contraction between tensors $T^{[S_1]}$ and $T^{[S_2]}$, can be computed as
\begin{align}
\overline{\#}{T^{[S_1,S_2]}} = \frac{\overline{\#}{T^{[S_1]}}\overline{\#}{T^{[S_2]}}}{|e_{S_1,S_2}|}.
\label{eq:norm_identity}
\end{align}
This follows from the fact that the leg represented by the edge $e_{S_1, S_2}$ does not matter for either the size or the capped size of $T^{[S_1,S_2]}$.

\begin{algorithm*}
  \caption{\texttt{LinDP}~\cite{adaptive}, adapted for tensor networks}\label{algo:lindp}
  \begin{algorithmic}[1]
  \STATE \textbf{Input:} Tensor network $\mathcal{T} = (V, E, c)$
  \STATE \textbf{Output:} General contraction tree $\tau^*$
    
  \STATE $\tau^* \leftarrow \varnothing$
  \FOR{\textbf{each} $T^{[i]}$ in $V$}
      \STATE $S \leftarrow$ Linearization of precedence graph $\mathcal{P}_{T^{[i]}}$ cf. \texttt{TensorIKKBZ} (Alg.~\ref{algo:tensor_ikkbz})
      \STATE Call \textsc{ChainDynamicProgramming}($S$) to build the optimal contraction tree $\tau$ respecting the linear order $S$
      \IF {$\tau^* = \varnothing$ or $\C(\tau) < \C(\tau^*)$}
            \STATE $\tau^* \leftarrow \tau$
      \ENDIF
  \ENDFOR
  \RETURN $\tau^*$
  \end{algorithmic}
\end{algorithm*}

The normalization step (Alg.~\ref{algo:normalization}) requires resolving contradictory sequences within a linearization. This is the only operation where we need to update scores. It receives a tensor $T$ (which will always be tensor $T^{[u]}$ from Alg.~\ref{algo:tensor_ikkbz}) along with the linearization $\mathcal{L}$ of its subtree, and performs multiple normalization steps until there are no contradictory sequences left. At each iteration, the score $\sigma(T)$ must be updated: the new numerator $\sigma_1(T)$ will represent the cost $\C(T)$ after performing the contraction with the tensor $\mathcal{L}_i$, the $i$th (compound) tensor in the linearization. To this end, we use Eq.~\eqref{eq:other_cout} to calculate the new cost (line~5). For the new denominator $\sigma_2(T)$, we calculate the capped size of the newly created tensor via Eq.~\eqref{eq:norm_identity} (line~6). Finally, we create the compound tensor which comprises both $T$ and $\mathcal{L}_i$ and replaces $T$ in the precedence graph (line~7). 
An execution of the \texttt{TensorIKKBZ} algorithm is illustrated in Appendix~\ref{appendix:tensor_ikkbz_example}.


\subsubsection{Time Complexity}
The first phase, constructing the precedence graph, can be done by depth-first search in linear time~\cite{dfs}, starting from the root and routing the edges to the unexplored nodes of the tree. In the second phase, linearization is the most expensive operation: Interleaving $m$ linearizations of total size $n - 1$ runs in $\mathcal{O}(n\log m)$-time via a min-heap data structure of size $m$~\cite{cormen}. In contrast, the normalization operation takes linear time. Since the algorithm repeats the two phases for each of the $n$ tensors, its total time complexity is $\mathcal{O}(n^2\log n)$~\cite{ik}.

Krishnamurthy et al.~\cite{kbz} improved the original \texttt{IKKBZ} algorithm to run in $\mathcal{O}(n^2)$-time by observing that the linearizations of two adjacent nodes are similar. We can apply the same observation in our context so that \texttt{TensorIKKBZ} runs in $\mathcal{O}(n^2)$-time as well.


\section{Generalized Optimization Approaches}\label{sec:beyond}

In the last section, we identified one of the settings under which we can provide optimal contraction orders. This was the case for \emph{tree} tensor networks and \emph{linear} contraction orders. In this section, we aim to close the gap to \emph{general} contraction orders and \emph{generic} tensor networks. Since the problem is NP-hard, we aim for a polynomial-time algorithm that yields a near-optimal solution.

Inspired by recent research in join ordering, we extend the previous algorithm to find near-optimal contraction trees respecting an initial \emph{fixed} permutation of the tensors. Recently, Ibrahim et al.~\cite{cubic_dp_in_tensor} have proposed a similar idea, however, they resort to heuristics to find the initial permutation. In contrast, we have already presented a way to order tensors: the \texttt{TensorIKKBZ} algorithm. Therefore, we can use its linear contraction orders as the initial permutation. This ensures that the optimality guarantee of the \texttt{TensorIKKBZ} algorithm is inherited for the case of a tree tensor network.

\subsection{Linearized Dynamic Programming}\label{subsec:lindp}

Textbooks introduce dynamic programming with the classical problem of matrix-chain multiplication and its cubic-time solution~\cite{cormen}. Matrices are a special class of tensors. Consequently, this problem is in fact the optimization problem on \emph{chain} tensor networks. To apply the same algorithm, we need to find a way to transform the tensor network into a chain, i.e., to find an order (permutation) of its tensors. We can then apply the cubic-time dynamic program to optimize the parenthesization of the tensors, which corresponds to the optimal contraction tree respecting the fixed permutation.

This idea was explored by Neumann and Radke~\cite{adaptive} in the context of join ordering. The novelty in their work consists in using the linearizations of the \texttt{IKKBZ} algorithm as seeds for the permutation, which guarantees optimality in the regime of linear orders. The technique is called linearized dynamic programming, \texttt{LinDP} for short, and has been shown to yield near-optimal \emph{general} join orders. We show that \texttt{LinDP} can be used with \texttt{TensorIKKBZ} to provide near-optimal \emph{general} contraction orders for tree tensor networks.

\textbf{\normalfont\bfseries Pseudocode.}\quad We outline the pseudocode in Alg.~\ref{algo:lindp} (see Appendix~\ref{appendix:lindp} for an extended version). In principle, one can apply the dynamic programming step only to the optimal linear order provided by \texttt{TensorIKKBZ}. Note, however, that \texttt{TensorIKKBZ} considers \emph{each} tensor as the root of the precedence graph anyway. As a consequence, we obtain $n$ many such linearizations, all of which can be used as initial permutations (line 5). This gives the algorithm more flexibility to find better general contraction orders.

\textbf{\normalfont\bfseries Time Complexity.}\quad The fact that \texttt{LinDP} considers all \texttt{TensorIKKBZ} linearizations makes it an expensive algorithm. Namely, its time complexity is $\mathcal{O}(n^4)$, since it runs the cubic-time dynamic program for each tensor separately (line 6). This problem can be alleviated by parallelizing the main for-loop (line~4).


\subsection{Generic Tensor Networks}\label{subsec:gtn}

While \texttt{LinDP} is also applicable to generic tensor networks, i.e., not necessarily trees, \texttt{TensorIKKBZ} requires as input a tree tensor network. As such, we need a mechanism to transform the input tensor network into a tree tensor network. A simple solution is to consider a \emph{spanning tree} of the network. This is a join ordering technique introduced by Krishnamurthy et al.~\cite{kbz} to deal with generic query graphs. In the experiments, we choose the \emph{maximum} spanning tree. The reason for this heuristic is as follows: In query optimization, the \emph{minimum} spanning tree is used (this follows from the observation that edges with low join selectivity are much more likely to be selected in the optimal plan). Given the correspondence between tensor contraction and the relational join that we will introduce in the next section, the choice of the maximum spanning tree is a natural adaptation of the original heuristic to tensor networks.


\section{Query Optimization}\label{sec:query_optimization}

As pointed out in the introduction, the new results are only possible due to a connection to join ordering, a subfield of database query optimization. Database systems provide a mechanism for storing tables and answering queries through a programming language called SQL (Structured Query Language)~\cite{sql}. The underlying mathematical foundation is relational algebra, which defines operators on tables (relations)~\cite{codd1970relational}.


\subsection{Query Graph}\label{subsec:query_graph}

\begin{figure}
    \centering
    \includegraphics{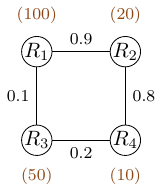}
    \caption[Query Graph]{Query graph with four tables. The vertex labels represent the table cardinalities and the edge weights represent the corresponding join selectivities.}
    \label{fig:query_graph}
\end{figure}

We are only interested in the binary relational operator \texttt{Join} ($\Join$), which takes two tables $A$ and $B$ as input and outputs another table $C = A\Join B$. A query contains multiple joins and is therefore graphically represented as a \emph{query graph}, where the vertices represent the tables and the edges represent the joins between them. The particularities of a query graph are discussed in detail in Appendix~\ref{appendix:query_graph}. In the sequel, we only need the notion of \emph{join selectivity}, defined as
\begin{equation}
f_{A,B} = \frac{|A \Join B|}{|A||B|},
\label{eq:join_sel}
\end{equation}
where $|\cdot|$ denotes the cardinality of the table, i.e., the number of records in the table. The join selectivity is used as the weight of the join edge between the two corresponding vertices in the query graph. A query graph with four tables along with their cardinalities is shown in Fig.~\ref{fig:query_graph}. For example, the cardinality of $R_1$ is 100 and the join selectivity of $R_1 \Join R_2$ in the same query graph is 0.9, i.e., $|R_1 \Join R_2| = 0.9\cdot 100 \cdot 20 = 1800$.

\subsection{Join Ordering}\label{subsec:join_ordering}

The result of a query consists of a single table that represents the join of all tables present in the query. However, joining the tables in the order specified by the user, e.g., by means of an SQL query~\cite{sql}, can lead to slow execution times. Therefore, the goal of query optimization and especially its subfield, join ordering, is to find or approximate the optimal execution order of the joins.


\subsubsection{Commonalities with Contraction Ordering}\label{subsec:commonalities}

The reader can already notice the similarities between our problem and the one present in join ordering. Indeed, the problems are almost identical: we can interpret a tensor contraction as a relational join. In the case where tables $A$ and $B$ do not share a common column, the relational join becomes a relational \emph{cross product}. This corresponds to the outer tensor product, i.e., there is no common leg to contract the two tensors over. Ideally, one should aim for solutions that can contain outer products, as this allows more flexibility. However, this comes at the cost of an exponentially larger search space during optimization.

\begin{figure}
    \centering
    \begin{subfigure}[b]{0.4\columnwidth}
        \centering
        \includegraphics[]{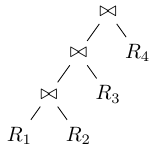}
        \caption{Left-deep (\emph{linear})}
    \end{subfigure}
    \begin{subfigure}[b]{0.4\columnwidth}
        \centering
        \includegraphics[]{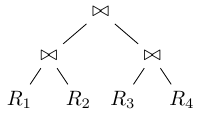}
        \caption{Bushy (\emph{general})}
    \end{subfigure}
    \caption[Join Tree Types]{Join tree types}
    \label{fig:join_tree_shapes}
\end{figure}

The equivalent of a contraction tree (Sec.~\ref{subsec:contraction_tree}) in query optimization is a \emph{join tree}, a rooted binary tree with tables as leaf nodes and join operators as inner nodes. The established classes of join trees are left-deep and bushy, illustrated in Fig.~\ref{fig:join_tree_shapes}. As the figure shows, they correspond to \emph{linear} and \emph{general} contraction trees, respectively, so we will only present the results in terms of \emph{linear} and \emph{general}.

\subsubsection{New Aspects of Contraction Ordering}\label{subsec:discrepancies}

There are several differences which, at first sight, impede results in join ordering, notably optimality results, to be applicable to tensor networks:


1) \emph{Tensor Contraction Cost.}\quad The sizes of the to-be-contracted legs between two tensors do not appear in the later contraction costs in which the contracted tensor is involved. In contrast, in join ordering, the weights of the join edges, which represent join selectivities, are still multiplied by in later join costs. An example is provided in Appendix~\ref{appendix:query_graph}.

2) \emph{Open Legs.}\quad Another particularity of tensor networks is the presence of open legs. Graphically, they are edges with only one endpoint and persist in the tensor network during its contraction. There is no explicit definition of them in the setting of join ordering we are considering, and thus the algorithms are not designed to support them.

The above differences are accounted for in Thm.~\ref{theorem:asi_cost_function} and the \texttt{TensorIKKBZ} algorithm. A key ingredient has been the reformulation of the tensor contraction cost in Eq.~\eqref{eqn:reformulation}, which is similar to the join cardinality in Eq.~\eqref{eq:join_sel}. Indeed, the factor $1 / \prod_{j=1}^{b} q_j$ in Eq.~\eqref{eqn:reformulation} can be interpreted as a join selectivity.


\section{Benchmarking and Evaluation}\label{sec:evaluation}

\begin{figure*}
    \centering
    \includegraphics[width=1.0\textwidth]{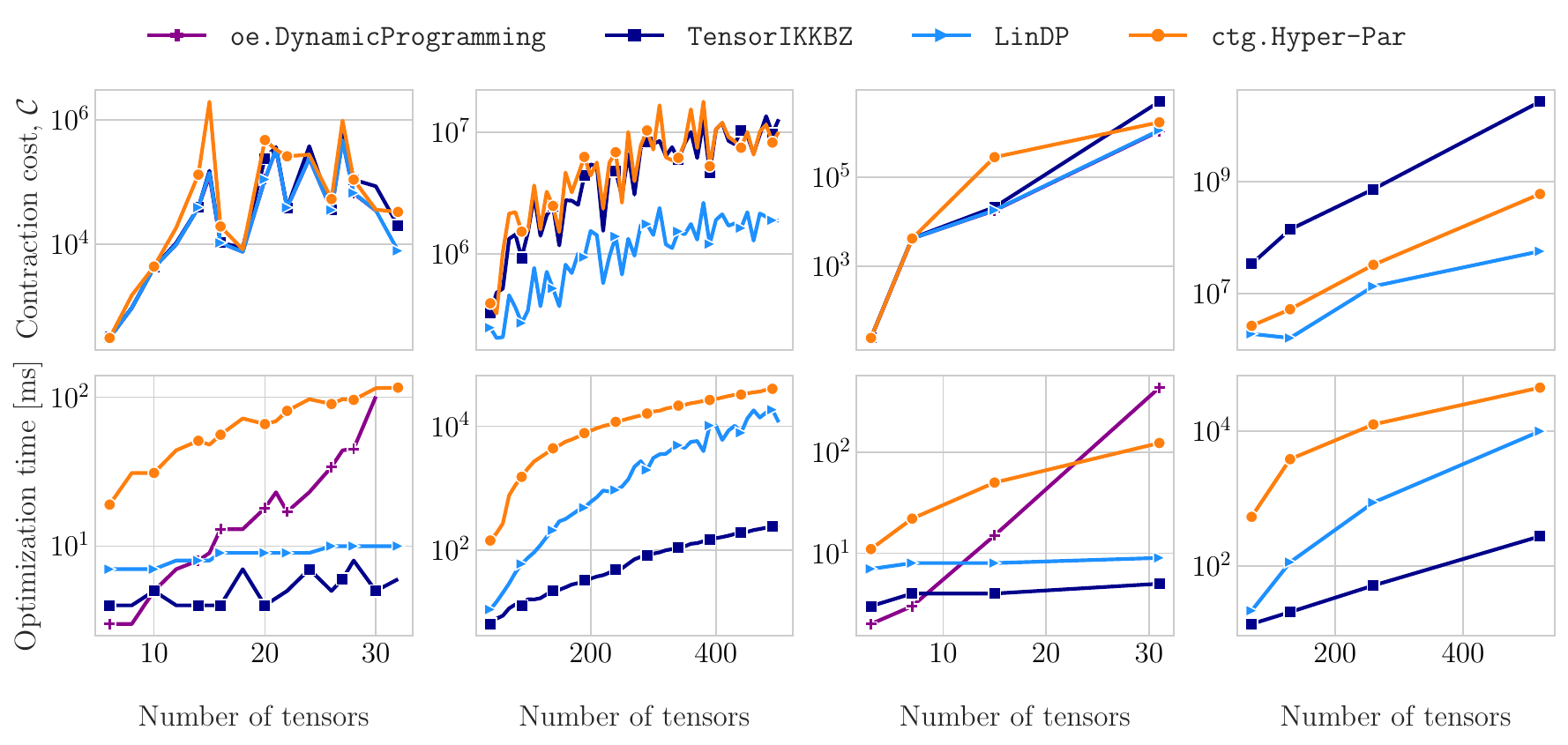}
    \begin{subfigure}{.5\textwidth}
        \centering
        \caption{Fork-Tensor Product State (FTPS)}
        \label{fig:myfirstsubfig}
    \end{subfigure}%
    \begin{subfigure}{.5\textwidth}
        \centering
        \caption{Hierarchical Tucker (HT)}
        \label{fig:mysecondsubfig}
    \end{subfigure}
    \caption{Benchmark of the time to find an optimized contraction scheme, and the resulting contraction cost for closed tree tensor networks. In both cases, lower is better.}
    \label{fig:eval}
\end{figure*}

To demonstrate the advantage of having an optimal algorithm as a starting point for further heuristics, we compare both \texttt{TensorIKKBZ} and \texttt{LinDP} against the recently proposed hyperoptimizer by Gray and Kourtis~\cite{cotengra}, \texttt{ctg.Hyper-Par}, which uses as subroutine the \texttt{KaHyPar} hypergraph partitioner~\cite{kahypar}. It has been shown to achieve competitive contraction costs on the tensor networks of several quantum circuits, including the Google Sycamore quantum circuit~\cite{sycamore}.

\textbf{\normalfont\bfseries Setting.}\quad We benchmark against two classes of tree tensor networks, the Hierarchical Tucker (HT) and the recently proposed Fork-Tensor Product State (FTPS)~\cite{ftps}. Their graphical notations are both illustrated in Appendix~\ref{appendix:graphical}. The leg sizes  are randomly chosen between 1 and 256. We also benchmark on the Sycamore circuit for which we need the spanning tree heuristic described in Sec.~\ref{subsec:gtn} (experiment results in Appendix~\ref{appendix:sycamore}).

In Fig.~\ref{fig:eval}, we plot the contraction cost, $\mathcal{C}$, and the optimization time for their \emph{closed} counterparts, i.e., without open legs. In the experiments, we let the algorithms fully exploit the potential of parallelization (in our case, 20 physical cores). Moreover, to ensure fairness, we let \texttt{ctg.Hyper-Par} sample contraction orders $n$ times (the default value is 128), since both \texttt{TensorIKKBZ} and \texttt{LinDP} generate $n$ orders as well. Both of our optimizers are implemented in \texttt{C++} and wrapped to be used directly in \texttt{opt\_einsum}~\cite{opt_einsum} and \texttt{cotengra}~\cite{cotengra}. Our implementation is publicly available.\footnote{\texttt{https://github.com/stoianmihail/Netzwerk}}

\textbf{\normalfont\bfseries Contraction Cost.}\quad The first observation is that, for FTPS, while linear contraction orders are more restrictive than general contraction orders, \texttt{TensorIKKBZ} is on par with \texttt{ctg.Hyper-Par}, which outputs general contraction orders. This is because \texttt{TensorIKKBZ} returns the optimal linear contraction order, thus providing robustness.

In addition, \texttt{LinDP} consistently outperforms \texttt{ctg.Hyper-Par} by almost an order of magnitude in both classes since it builds the optimal contraction tree respecting \texttt{TensorIKKBZ}'s linearization. For small instances, it is on par with the exact exponential dynamic programming of \texttt{opt\_einsum}, \texttt{oe.DynamicProgramming}.

\textbf{\normalfont\bfseries Optimization Time.}\quad Both of our optimizers are faster than \texttt{ctg.Hyper-Par}. This is partly due to the performance of \texttt{C++}, but also to the good time complexity of the algorithms and their parallelization potential.


\section{Conclusion}\label{sec:conclusions}

We proved that tree tensor networks admit optimal linear contraction orders in polynomial time. To this end, we provided the \texttt{TensorIKKBZ} algorithm, which is an adaptation of a decades-old algorithm from database join ordering. It requires the cost function to satisfy the adjacent sequence interchange (ASI) property. Thus, we proved that the cost function used in the context of tensor networks does indeed satisfy this property.

Apart from the optimality result, we have employed \texttt{TensorIKKBZ} in \texttt{LinDP}~\cite{adaptive}, a recent join ordering optimizer that outputs the optimal general contraction tree respecting an initial permutation of the tensors. Its advantage over previous work is that it inherits the optimality guarantees of the \texttt{TensorIKKBZ} algorithm. To cover the full optimization spectrum, we argued that selecting the maximum spanning tree is a well-suited heuristic for optimizing generic tensor networks.

\textbf{\normalfont\bfseries Future Work.}\quad In future work, one should attempt to support hyperedges in \texttt{TensorIKKBZ}, since there are several instances where hyperedges occur in tensor networks, such as in the 70-qubit Bristlecone~\cite{bristlecone}. While \texttt{LinDP} can easily support hyperedges, \texttt{TensorIKKBZ} is not trivially extendable. The work of Radke and Neumann~\cite{radke} in join ordering is not directly applicable since hyperedges in tensor networks are not binary. Furthermore, we assumed a naive implementation of tensor contraction which leads to the contraction cost in Eq.~\eqref{eq:tensor_contraction_cost}. Instead one could consider more efficient implementations, such as Strassen's algorithm~\cite{Huang2018}.


\section*{Acknowledgement}

We like to thank Altan Birler for his valuable feedback and proofreading of the sections on query optimization. The research is part of the Munich Quantum Valley, which is supported by the Bavarian state government with funds from the Hightech Agenda Bayern Plus.


\bibliography{bibliography}

\begin{thebibliography}{38}%
\makeatletter
\providecommand \@ifxundefined [1]{%
 \@ifx{#1\undefined}
}%
\providecommand \@ifnum [1]{%
 \ifnum #1\expandafter \@firstoftwo
 \else \expandafter \@secondoftwo
 \fi
}%
\providecommand \@ifx [1]{%
 \ifx #1\expandafter \@firstoftwo
 \else \expandafter \@secondoftwo
 \fi
}%
\providecommand \natexlab [1]{#1}%
\providecommand \enquote  [1]{``#1''}%
\providecommand \bibnamefont  [1]{#1}%
\providecommand \bibfnamefont [1]{#1}%
\providecommand \citenamefont [1]{#1}%
\providecommand \href@noop [0]{\@secondoftwo}%
\providecommand \href [0]{\begingroup \@sanitize@url \@href}%
\providecommand \@href[1]{\@@startlink{#1}\@@href}%
\providecommand \@@href[1]{\endgroup#1\@@endlink}%
\providecommand \@sanitize@url [0]{\catcode `\\12\catcode `\$12\catcode `\&12\catcode `\#12\catcode `\^12\catcode `\_12\catcode `\%12\relax}%
\providecommand \@@startlink[1]{}%
\providecommand \@@endlink[0]{}%
\providecommand \url  [0]{\begingroup\@sanitize@url \@url }%
\providecommand \@url [1]{\endgroup\@href {#1}{\urlprefix }}%
\providecommand \urlprefix  [0]{URL }%
\providecommand \Eprint [0]{\href }%
\providecommand \doibase [0]{https://doi.org/}%
\providecommand \selectlanguage [0]{\@gobble}%
\providecommand \bibinfo  [0]{\@secondoftwo}%
\providecommand \bibfield  [0]{\@secondoftwo}%
\providecommand \translation [1]{[#1]}%
\providecommand \BibitemOpen [0]{}%
\providecommand \bibitemStop [0]{}%
\providecommand \bibitemNoStop [0]{.\EOS\space}%
\providecommand \EOS [0]{\spacefactor3000\relax}%
\providecommand \BibitemShut  [1]{\csname bibitem#1\endcsname}%
\let\auto@bib@innerbib\@empty
\bibitem [{\citenamefont {Schollw\"ock}(2011)}]{Schollwoeck2011}%
  \BibitemOpen
  \bibfield  {author} {\bibinfo {author} {\bibfnamefont {U.}~\bibnamefont {Schollw\"ock}},\ }\bibfield  {title} {\bibinfo {title} {The density-matrix renormalization group in the age of matrix product states},\ }\href {https://doi.org/10.1016/j.aop.2010.09.012} {\bibfield  {journal} {\bibinfo  {journal} {Ann. Physics}\ }\textbf {\bibinfo {volume} {326}},\ \bibinfo {pages} {96} (\bibinfo {year} {2011})}\BibitemShut {NoStop}%
\bibitem [{\citenamefont {Hackbusch}(2012)}]{hackbusch}%
  \BibitemOpen
  \bibfield  {author} {\bibinfo {author} {\bibfnamefont {W.}~\bibnamefont {Hackbusch}},\ }\href {https://doi.org/10.1007/978-3-642-28027-6} {\emph {\bibinfo {title} {{Tensor spaces and numerical tensor calculus}}}},\ \bibinfo {series} {Springer series in computational mathematics}, Vol.~\bibinfo {volume} {42}\ (\bibinfo  {publisher} {Springer},\ \bibinfo {address} {Heidelberg},\ \bibinfo {year} {2012})\ pp.\ \bibinfo {pages} {xxiv, 500}\BibitemShut {NoStop}%
\bibitem [{\citenamefont {Cichocki}\ \emph {et~al.}(2016)\citenamefont {Cichocki}, \citenamefont {Lee}, \citenamefont {Oseledets}, \citenamefont {Phan}, \citenamefont {Zhao},\ and\ \citenamefont {Mandic}}]{MAL-059}%
  \BibitemOpen
  \bibfield  {author} {\bibinfo {author} {\bibfnamefont {A.}~\bibnamefont {Cichocki}}, \bibinfo {author} {\bibfnamefont {N.}~\bibnamefont {Lee}}, \bibinfo {author} {\bibfnamefont {I.}~\bibnamefont {Oseledets}}, \bibinfo {author} {\bibfnamefont {A.-H.}\ \bibnamefont {Phan}}, \bibinfo {author} {\bibfnamefont {Q.}~\bibnamefont {Zhao}},\ and\ \bibinfo {author} {\bibfnamefont {D.~P.}\ \bibnamefont {Mandic}},\ }\bibfield  {title} {\bibinfo {title} {Tensor networks for dimensionality reduction and large-scale optimization: {P}art 1 {L}ow-rank tensor decompositions},\ }\href {https://doi.org/10.1561/2200000059} {\bibfield  {journal} {\bibinfo  {journal} {Foundations and Trends in Machine Learning}\ }\textbf {\bibinfo {volume} {9}},\ \bibinfo {pages} {249} (\bibinfo {year} {2016})}\BibitemShut {NoStop}%
\bibitem [{\citenamefont {Cichocki}\ \emph {et~al.}(2017)\citenamefont {Cichocki}, \citenamefont {Phan}, \citenamefont {Zhao}, \citenamefont {Lee}, \citenamefont {Oseledets}, \citenamefont {Sugiyama},\ and\ \citenamefont {Mandic}}]{MAL-067}%
  \BibitemOpen
  \bibfield  {author} {\bibinfo {author} {\bibfnamefont {A.}~\bibnamefont {Cichocki}}, \bibinfo {author} {\bibfnamefont {A.-H.}\ \bibnamefont {Phan}}, \bibinfo {author} {\bibfnamefont {Q.}~\bibnamefont {Zhao}}, \bibinfo {author} {\bibfnamefont {N.}~\bibnamefont {Lee}}, \bibinfo {author} {\bibfnamefont {I.}~\bibnamefont {Oseledets}}, \bibinfo {author} {\bibfnamefont {M.}~\bibnamefont {Sugiyama}},\ and\ \bibinfo {author} {\bibfnamefont {D.~P.}\ \bibnamefont {Mandic}},\ }\bibfield  {title} {\bibinfo {title} {Tensor networks for dimensionality reduction and large-scale optimization: {P}art 2 {A}pplications and future perspectives},\ }\href {https://doi.org/10.1561/2200000067} {\bibfield  {journal} {\bibinfo  {journal} {Foundations and Trends in Machine Learning}\ }\textbf {\bibinfo {volume} {9}},\ \bibinfo {pages} {431} (\bibinfo {year} {2017})}\BibitemShut {NoStop}%
\bibitem [{\citenamefont {Markov}\ and\ \citenamefont {Shi}(2008{\natexlab{a}})}]{quantum_via_tn}%
  \BibitemOpen
  \bibfield  {author} {\bibinfo {author} {\bibfnamefont {I.~L.}\ \bibnamefont {Markov}}\ and\ \bibinfo {author} {\bibfnamefont {Y.}~\bibnamefont {Shi}},\ }\bibfield  {title} {\bibinfo {title} {Simulating quantum computation by contracting tensor networks},\ }\href {https://doi.org/10.1137/050644756} {\bibfield  {journal} {\bibinfo  {journal} {SIAM J. Comput.}\ }\textbf {\bibinfo {volume} {38}},\ \bibinfo {pages} {963} (\bibinfo {year} {2008}{\natexlab{a}})}\BibitemShut {NoStop}%
\bibitem [{\citenamefont {Stoudenmire}\ and\ \citenamefont {Schwab}(2016)}]{Stoudenmire2016SupervisedLW}%
  \BibitemOpen
  \bibfield  {author} {\bibinfo {author} {\bibfnamefont {E.~M.}\ \bibnamefont {Stoudenmire}}\ and\ \bibinfo {author} {\bibfnamefont {D.~J.}\ \bibnamefont {Schwab}},\ }\bibfield  {title} {\bibinfo {title} {Supervised learning with tensor networks},\ }in\ \href@noop {} {\emph {\bibinfo {booktitle} {Proceedings of the 30th International Conference on Neural Information Processing Systems}}},\ \bibinfo {series and number} {NIPS'16}\ (\bibinfo  {publisher} {Curran Associates Inc.},\ \bibinfo {address} {Red Hook, NY, USA},\ \bibinfo {year} {2016})\ pp.\ \bibinfo {pages} {4806--4814}\BibitemShut {NoStop}%
\bibitem [{\citenamefont {Szalay}\ \emph {et~al.}(2015)\citenamefont {Szalay}, \citenamefont {Pfeffer}, \citenamefont {Murg}, \citenamefont {Barcza}, \citenamefont {Verstraete}, \citenamefont {Schneider},\ and\ \citenamefont {Legeza}}]{Szalay2015}%
  \BibitemOpen
  \bibfield  {author} {\bibinfo {author} {\bibfnamefont {S.}~\bibnamefont {Szalay}}, \bibinfo {author} {\bibfnamefont {M.}~\bibnamefont {Pfeffer}}, \bibinfo {author} {\bibfnamefont {V.}~\bibnamefont {Murg}}, \bibinfo {author} {\bibfnamefont {G.}~\bibnamefont {Barcza}}, \bibinfo {author} {\bibfnamefont {F.}~\bibnamefont {Verstraete}}, \bibinfo {author} {\bibfnamefont {R.}~\bibnamefont {Schneider}},\ and\ \bibinfo {author} {\bibfnamefont {O.}~\bibnamefont {Legeza}},\ }\bibfield  {title} {\bibinfo {title} {Tensor product methods and entanglement optimization for ab initio quantum chemistry},\ }\href {https://doi.org/10.1002/qua.24898} {\bibfield  {journal} {\bibinfo  {journal} {Int. J. Quantum Chem.}\ }\textbf {\bibinfo {volume} {115}},\ \bibinfo {pages} {1342} (\bibinfo {year} {2015})}\BibitemShut {NoStop}%
\bibitem [{\citenamefont {Abo~Khamis}\ \emph {et~al.}(2016)\citenamefont {Abo~Khamis}, \citenamefont {Ngo},\ and\ \citenamefont {Rudra}}]{khamis_faq}%
  \BibitemOpen
  \bibfield  {author} {\bibinfo {author} {\bibfnamefont {M.}~\bibnamefont {Abo~Khamis}}, \bibinfo {author} {\bibfnamefont {H.~Q.}\ \bibnamefont {Ngo}},\ and\ \bibinfo {author} {\bibfnamefont {A.}~\bibnamefont {Rudra}},\ }\bibfield  {title} {\bibinfo {title} {{FAQ}: {Q}uestions {A}sked {F}requently},\ }in\ \href {https://doi.org/10.1145/2902251.2902280} {\emph {\bibinfo {booktitle} {Proceedings of the 35th ACM SIGMOD-SIGACT-SIGAI Symposium on Principles of Database Systems}}},\ \bibinfo {series and number} {PODS '16}\ (\bibinfo  {publisher} {Association for Computing Machinery},\ \bibinfo {address} {New York, NY, USA},\ \bibinfo {year} {2016})\ pp.\ \bibinfo {pages} {13--28}\BibitemShut {NoStop}%
\bibitem [{\citenamefont {Dudek}\ \emph {et~al.}(2019)\citenamefont {Dudek}, \citenamefont {Due{\~{n}}as{-}Osorio},\ and\ \citenamefont {Vardi}}]{dudek2020efficient}%
  \BibitemOpen
  \bibfield  {author} {\bibinfo {author} {\bibfnamefont {J.~M.}\ \bibnamefont {Dudek}}, \bibinfo {author} {\bibfnamefont {L.}~\bibnamefont {Due{\~{n}}as{-}Osorio}},\ and\ \bibinfo {author} {\bibfnamefont {M.~Y.}\ \bibnamefont {Vardi}},\ }\bibfield  {title} {\bibinfo {title} {Efficient contraction of large tensor networks for weighted model counting through graph decompositions},\ }\href {http://arxiv.org/abs/1908.04381} {\bibfield  {journal} {\bibinfo  {journal} {CoRR}\ }\textbf {\bibinfo {volume} {abs/1908.04381}} (\bibinfo {year} {2019})},\ \Eprint {https://arxiv.org/abs/1908.04381} {1908.04381} \BibitemShut {NoStop}%
\bibitem [{\citenamefont {Lam}\ \emph {et~al.}(1997)\citenamefont {Lam}, \citenamefont {Sadayappan},\ and\ \citenamefont {Wenger}}]{Lam1997OnOA}%
  \BibitemOpen
  \bibfield  {author} {\bibinfo {author} {\bibfnamefont {C.-C.}\ \bibnamefont {Lam}}, \bibinfo {author} {\bibfnamefont {P.}~\bibnamefont {Sadayappan}},\ and\ \bibinfo {author} {\bibfnamefont {R.}~\bibnamefont {Wenger}},\ }\bibfield  {title} {\bibinfo {title} {On optimizing a class of multi-dimensional loops with reductions for parallel execution},\ }\href {https://doi.org/10.1142/S0129626497000176} {\bibfield  {journal} {\bibinfo  {journal} {Parallel Process. Lett.}\ }\textbf {\bibinfo {volume} {7}},\ \bibinfo {pages} {157} (\bibinfo {year} {1997})}\BibitemShut {NoStop}%
\bibitem [{\citenamefont {Pfeifer}\ \emph {et~al.}(2014)\citenamefont {Pfeifer}, \citenamefont {Haegeman},\ and\ \citenamefont {Verstraete}}]{dp_in_tensor_network}%
  \BibitemOpen
  \bibfield  {author} {\bibinfo {author} {\bibfnamefont {R.~N.~C.}\ \bibnamefont {Pfeifer}}, \bibinfo {author} {\bibfnamefont {J.}~\bibnamefont {Haegeman}},\ and\ \bibinfo {author} {\bibfnamefont {F.}~\bibnamefont {Verstraete}},\ }\bibfield  {title} {\bibinfo {title} {Faster identification of optimal contraction sequences for tensor networks},\ }\href {https://doi.org/10.1103/PhysRevE.90.033315} {\bibfield  {journal} {\bibinfo  {journal} {Phys. Rev. E}\ }\textbf {\bibinfo {volume} {90}},\ \bibinfo {pages} {033315} (\bibinfo {year} {2014})}\BibitemShut {NoStop}%
\bibitem [{\citenamefont {Gray}\ and\ \citenamefont {Kourtis}(2021)}]{cotengra}%
  \BibitemOpen
  \bibfield  {author} {\bibinfo {author} {\bibfnamefont {J.}~\bibnamefont {Gray}}\ and\ \bibinfo {author} {\bibfnamefont {S.}~\bibnamefont {Kourtis}},\ }\bibfield  {title} {\bibinfo {title} {Hyper-optimized tensor network contraction},\ }\href {https://doi.org/10.22331/q-2021-03-15-410} {\bibfield  {journal} {\bibinfo  {journal} {Quantum}\ }\textbf {\bibinfo {volume} {5}},\ \bibinfo {pages} {410} (\bibinfo {year} {2021})}\BibitemShut {NoStop}%
\bibitem [{\citenamefont {Chen}\ \emph {et~al.}(2018)\citenamefont {Chen}, \citenamefont {Zhang}, \citenamefont {Huang}, \citenamefont {Newman},\ and\ \citenamefont {Shi}}]{chen2018classical}%
  \BibitemOpen
  \bibfield  {author} {\bibinfo {author} {\bibfnamefont {J.}~\bibnamefont {Chen}}, \bibinfo {author} {\bibfnamefont {F.}~\bibnamefont {Zhang}}, \bibinfo {author} {\bibfnamefont {C.}~\bibnamefont {Huang}}, \bibinfo {author} {\bibfnamefont {M.}~\bibnamefont {Newman}},\ and\ \bibinfo {author} {\bibfnamefont {Y.}~\bibnamefont {Shi}},\ }\bibfield  {title} {\bibinfo {title} {Classical simulation of intermediate-size quantum circuits},\ }\href@noop {} {\  (\bibinfo {year} {2018})},\ \Eprint {https://arxiv.org/abs/1805.01450} {arXiv:1805.01450 [quant-ph]} \BibitemShut {NoStop}%
\bibitem [{\citenamefont {Huang}\ \emph {et~al.}(2020)\citenamefont {Huang}, \citenamefont {Zhang}, \citenamefont {Newman}, \citenamefont {Cai}, \citenamefont {Gao}, \citenamefont {Tian}, \citenamefont {Wu}, \citenamefont {Xu}, \citenamefont {Yu}, \citenamefont {Yuan}, \citenamefont {Szegedy}, \citenamefont {Shi},\ and\ \citenamefont {Chen}}]{huang2020classical}%
  \BibitemOpen
  \bibfield  {author} {\bibinfo {author} {\bibfnamefont {C.}~\bibnamefont {Huang}}, \bibinfo {author} {\bibfnamefont {F.}~\bibnamefont {Zhang}}, \bibinfo {author} {\bibfnamefont {M.}~\bibnamefont {Newman}}, \bibinfo {author} {\bibfnamefont {J.}~\bibnamefont {Cai}}, \bibinfo {author} {\bibfnamefont {X.}~\bibnamefont {Gao}}, \bibinfo {author} {\bibfnamefont {Z.}~\bibnamefont {Tian}}, \bibinfo {author} {\bibfnamefont {J.}~\bibnamefont {Wu}}, \bibinfo {author} {\bibfnamefont {H.}~\bibnamefont {Xu}}, \bibinfo {author} {\bibfnamefont {H.}~\bibnamefont {Yu}}, \bibinfo {author} {\bibfnamefont {B.}~\bibnamefont {Yuan}}, \bibinfo {author} {\bibfnamefont {M.}~\bibnamefont {Szegedy}}, \bibinfo {author} {\bibfnamefont {Y.}~\bibnamefont {Shi}},\ and\ \bibinfo {author} {\bibfnamefont {J.}~\bibnamefont {Chen}},\ }\bibfield  {title} {\bibinfo {title} {Classical simulation of quantum supremacy circuits},\ }\href@noop {} {\  (\bibinfo {year} {2020})},\ \Eprint {https://arxiv.org/abs/2005.06787} {arXiv:2005.06787 [quant-ph]}
  \BibitemShut {NoStop}%
\bibitem [{\citenamefont {Xu}\ \emph {et~al.}(2019)\citenamefont {Xu}, \citenamefont {Liang}, \citenamefont {Deng}, \citenamefont {Wen}, \citenamefont {Xie},\ and\ \citenamefont {Li}}]{ttn_optimal_max}%
  \BibitemOpen
  \bibfield  {author} {\bibinfo {author} {\bibfnamefont {J.}~\bibnamefont {Xu}}, \bibinfo {author} {\bibfnamefont {L.}~\bibnamefont {Liang}}, \bibinfo {author} {\bibfnamefont {L.}~\bibnamefont {Deng}}, \bibinfo {author} {\bibfnamefont {C.}~\bibnamefont {Wen}}, \bibinfo {author} {\bibfnamefont {Y.}~\bibnamefont {Xie}},\ and\ \bibinfo {author} {\bibfnamefont {G.}~\bibnamefont {Li}},\ }\bibfield  {title} {\bibinfo {title} {Towards a polynomial algorithm for optimal contraction sequence of tensor networks from trees},\ }\href {https://doi.org/10.1103/PhysRevE.100.043309} {\bibfield  {journal} {\bibinfo  {journal} {Phys. Rev. E}\ }\textbf {\bibinfo {volume} {100}},\ \bibinfo {pages} {043309} (\bibinfo {year} {2019})}\BibitemShut {NoStop}%
\bibitem [{\citenamefont {Ibrahim}\ \emph {et~al.}(2022)\citenamefont {Ibrahim}, \citenamefont {Lykov}, \citenamefont {He}, \citenamefont {Alexeev},\ and\ \citenamefont {Safro}}]{cubic_dp_in_tensor}%
  \BibitemOpen
  \bibfield  {author} {\bibinfo {author} {\bibfnamefont {C.}~\bibnamefont {Ibrahim}}, \bibinfo {author} {\bibfnamefont {D.}~\bibnamefont {Lykov}}, \bibinfo {author} {\bibfnamefont {Z.}~\bibnamefont {He}}, \bibinfo {author} {\bibfnamefont {Y.}~\bibnamefont {Alexeev}},\ and\ \bibinfo {author} {\bibfnamefont {I.}~\bibnamefont {Safro}},\ }\bibfield  {title} {\bibinfo {title} {Constructing optimal contraction trees for tensor network quantum circuit simulation},\ }in\ \href {https://doi.org/10.1109/HPEC55821.2022.9926353} {\emph {\bibinfo {booktitle} {2022 IEEE High Performance Extreme Computing Conference (HPEC)}}}\ (\bibinfo {year} {2022})\ pp.\ \bibinfo {pages} {1--8}\BibitemShut {NoStop}%
\bibitem [{\citenamefont {Cormen}\ \emph {et~al.}(2009)\citenamefont {Cormen}, \citenamefont {Leiserson}, \citenamefont {Rivest},\ and\ \citenamefont {Stein}}]{cormen}%
  \BibitemOpen
  \bibfield  {author} {\bibinfo {author} {\bibfnamefont {T.~H.}\ \bibnamefont {Cormen}}, \bibinfo {author} {\bibfnamefont {C.~E.}\ \bibnamefont {Leiserson}}, \bibinfo {author} {\bibfnamefont {R.~L.}\ \bibnamefont {Rivest}},\ and\ \bibinfo {author} {\bibfnamefont {C.}~\bibnamefont {Stein}},\ }\href@noop {} {\emph {\bibinfo {title} {Introduction to Algorithms, Third Edition}}},\ \bibinfo {edition} {3rd}\ ed.\ (\bibinfo  {publisher} {The MIT Press},\ \bibinfo {year} {2009})\BibitemShut {NoStop}%
\bibitem [{\citenamefont {Neumann}\ and\ \citenamefont {Radke}(2018)}]{adaptive}%
  \BibitemOpen
  \bibfield  {author} {\bibinfo {author} {\bibfnamefont {T.}~\bibnamefont {Neumann}}\ and\ \bibinfo {author} {\bibfnamefont {B.}~\bibnamefont {Radke}},\ }\bibfield  {title} {\bibinfo {title} {Adaptive optimization of very large join queries},\ }in\ \href {https://doi.org/10.1145/3183713.3183733} {\emph {\bibinfo {booktitle} {Proceedings of the 2018 International Conference on Management of Data}}},\ \bibinfo {series and number} {SIGMOD '18}\ (\bibinfo  {publisher} {Association for Computing Machinery},\ \bibinfo {address} {New York, NY, USA},\ \bibinfo {year} {2018})\ pp.\ \bibinfo {pages} {677--692}\BibitemShut {NoStop}%
\bibitem [{\citenamefont {Markov}\ and\ \citenamefont {Shi}(2008{\natexlab{b}})}]{shi_tree_decomp}%
  \BibitemOpen
  \bibfield  {author} {\bibinfo {author} {\bibfnamefont {I.~L.}\ \bibnamefont {Markov}}\ and\ \bibinfo {author} {\bibfnamefont {Y.}~\bibnamefont {Shi}},\ }\bibfield  {title} {\bibinfo {title} {Simulating quantum computation by contracting tensor networks},\ }\href {https://doi.org/10.1137/050644756} {\bibfield  {journal} {\bibinfo  {journal} {SIAM Journal on Computing}\ }\textbf {\bibinfo {volume} {38}},\ \bibinfo {pages} {963} (\bibinfo {year} {2008}{\natexlab{b}})},\ \Eprint {https://arxiv.org/abs/https://doi.org/10.1137/050644756} {https://doi.org/10.1137/050644756} \BibitemShut {NoStop}%
\bibitem [{\citenamefont {Cygan}\ \emph {et~al.}(2015)\citenamefont {Cygan}, \citenamefont {Fomin}, \citenamefont {Kowalik}, \citenamefont {Lokshtanov}, \citenamefont {Marx}, \citenamefont {Pilipczuk}, \citenamefont {Pilipczuk},\ and\ \citenamefont {Saurabh}}]{Cygan_book}%
  \BibitemOpen
  \bibfield  {author} {\bibinfo {author} {\bibfnamefont {M.}~\bibnamefont {Cygan}}, \bibinfo {author} {\bibfnamefont {F.~V.}\ \bibnamefont {Fomin}}, \bibinfo {author} {\bibfnamefont {L.}~\bibnamefont {Kowalik}}, \bibinfo {author} {\bibfnamefont {D.}~\bibnamefont {Lokshtanov}}, \bibinfo {author} {\bibfnamefont {D.}~\bibnamefont {Marx}}, \bibinfo {author} {\bibfnamefont {M.}~\bibnamefont {Pilipczuk}}, \bibinfo {author} {\bibfnamefont {M.}~\bibnamefont {Pilipczuk}},\ and\ \bibinfo {author} {\bibfnamefont {S.}~\bibnamefont {Saurabh}},\ }\href@noop {} {\emph {\bibinfo {title} {Parameterized Algorithms}}},\ \bibinfo {edition} {1st}\ ed.\ (\bibinfo  {publisher} {Springer Publishing Company, Incorporated},\ \bibinfo {year} {2015})\BibitemShut {NoStop}%
\bibitem [{\citenamefont {Boixo}\ \emph {et~al.}(2018{\natexlab{a}})\citenamefont {Boixo}, \citenamefont {Isakov}, \citenamefont {Smelyanskiy},\ and\ \citenamefont {Neven}}]{boixo2018simulation}%
  \BibitemOpen
  \bibfield  {author} {\bibinfo {author} {\bibfnamefont {S.}~\bibnamefont {Boixo}}, \bibinfo {author} {\bibfnamefont {S.~V.}\ \bibnamefont {Isakov}}, \bibinfo {author} {\bibfnamefont {V.~N.}\ \bibnamefont {Smelyanskiy}},\ and\ \bibinfo {author} {\bibfnamefont {H.}~\bibnamefont {Neven}},\ }\href@noop {} {\bibinfo {title} {Simulation of low-depth quantum circuits as complex undirected graphical models}} (\bibinfo {year} {2018}{\natexlab{a}}),\ \Eprint {https://arxiv.org/abs/1712.05384} {arXiv:1712.05384 [quant-ph]} \BibitemShut {NoStop}%
\bibitem [{\citenamefont {Schutski}\ \emph {et~al.}(2020)\citenamefont {Schutski}, \citenamefont {Lykov},\ and\ \citenamefont {Oseledets}}]{roman_tree_decomp}%
  \BibitemOpen
  \bibfield  {author} {\bibinfo {author} {\bibfnamefont {R.}~\bibnamefont {Schutski}}, \bibinfo {author} {\bibfnamefont {D.}~\bibnamefont {Lykov}},\ and\ \bibinfo {author} {\bibfnamefont {I.}~\bibnamefont {Oseledets}},\ }\bibfield  {title} {\bibinfo {title} {Adaptive algorithm for quantum circuit simulation},\ }\href {https://doi.org/10.1103/PhysRevA.101.042335} {\bibfield  {journal} {\bibinfo  {journal} {Phys. Rev. A}\ }\textbf {\bibinfo {volume} {101}},\ \bibinfo {pages} {042335} (\bibinfo {year} {2020})}\BibitemShut {NoStop}%
\bibitem [{\citenamefont {O'Gorman}(2019)}]{linear_and_general}%
  \BibitemOpen
  \bibfield  {author} {\bibinfo {author} {\bibfnamefont {B.}~\bibnamefont {O'Gorman}},\ }\bibfield  {title} {\bibinfo {title} {Parameterization of tensor network contraction},\ }in\ \href {https://doi.org/10.4230/LIPIcs.TQC.2019.10} {\emph {\bibinfo {booktitle} {14th Conference on the Theory of Quantum Computation, Communication and Cryptography (TQC 2019)}}},\ \bibinfo {series} {Leibniz International Proceedings in Informatics (LIPIcs)}, Vol.\ \bibinfo {volume} {135},\ \bibinfo {editor} {edited by\ \bibinfo {editor} {\bibfnamefont {W.}~\bibnamefont {van Dam}}\ and\ \bibinfo {editor} {\bibfnamefont {L.}~\bibnamefont {Mancinska}}}\ (\bibinfo  {publisher} {Schloss Dagstuhl--Leibniz-Zentrum fuer Informatik},\ \bibinfo {address} {Dagstuhl, Germany},\ \bibinfo {year} {2019})\ pp.\ \bibinfo {pages} {10:1--10:19}\BibitemShut {NoStop}%
\bibitem [{\citenamefont {Evenbly}\ and\ \citenamefont {Pfeifer}(2014)}]{contraction_tree_first}%
  \BibitemOpen
  \bibfield  {author} {\bibinfo {author} {\bibfnamefont {G.}~\bibnamefont {Evenbly}}\ and\ \bibinfo {author} {\bibfnamefont {R.~N.~C.}\ \bibnamefont {Pfeifer}},\ }\bibfield  {title} {\bibinfo {title} {Improving the efficiency of variational tensor network algorithms},\ }\href {https://doi.org/10.1103/PhysRevB.89.245118} {\bibfield  {journal} {\bibinfo  {journal} {Phys. Rev. B}\ }\textbf {\bibinfo {volume} {89}},\ \bibinfo {pages} {245118} (\bibinfo {year} {2014})}\BibitemShut {NoStop}%
\bibitem [{\citenamefont {Monma}\ and\ \citenamefont {Sidney}(1979)}]{Monma1979SequencingWS}%
  \BibitemOpen
  \bibfield  {author} {\bibinfo {author} {\bibfnamefont {C.~L.}\ \bibnamefont {Monma}}\ and\ \bibinfo {author} {\bibfnamefont {J.~B.}\ \bibnamefont {Sidney}},\ }\bibfield  {title} {\bibinfo {title} {Sequencing with series-parallel precedence constraints},\ }\href {http://www.jstor.org/stable/3689575} {\bibfield  {journal} {\bibinfo  {journal} {Mathematics of Operations Research}\ }\textbf {\bibinfo {volume} {4}},\ \bibinfo {pages} {215} (\bibinfo {year} {1979})}\BibitemShut {NoStop}%
\bibitem [{\citenamefont {Ibaraki}\ and\ \citenamefont {Kameda}(1984)}]{ik}%
  \BibitemOpen
  \bibfield  {author} {\bibinfo {author} {\bibfnamefont {T.}~\bibnamefont {Ibaraki}}\ and\ \bibinfo {author} {\bibfnamefont {T.}~\bibnamefont {Kameda}},\ }\bibfield  {title} {\bibinfo {title} {On the optimal nesting order for computing {N}-relational joins},\ }\href {https://doi.org/10.1145/1270.1498} {\bibfield  {journal} {\bibinfo  {journal} {ACM Trans. Database Syst.}\ }\textbf {\bibinfo {volume} {9}},\ \bibinfo {pages} {482} (\bibinfo {year} {1984})}\BibitemShut {NoStop}%
\bibitem [{\citenamefont {Krishnamurthy}\ \emph {et~al.}(1986)\citenamefont {Krishnamurthy}, \citenamefont {Boral},\ and\ \citenamefont {Zaniolo}}]{kbz}%
  \BibitemOpen
  \bibfield  {author} {\bibinfo {author} {\bibfnamefont {R.}~\bibnamefont {Krishnamurthy}}, \bibinfo {author} {\bibfnamefont {H.}~\bibnamefont {Boral}},\ and\ \bibinfo {author} {\bibfnamefont {C.}~\bibnamefont {Zaniolo}},\ }\bibfield  {title} {\bibinfo {title} {Optimization of nonrecursive queries},\ }in\ \href@noop {} {\emph {\bibinfo {booktitle} {Proceedings of the 12th International Conference on Very Large Data Bases}}},\ \bibinfo {series and number} {VLDB '86}\ (\bibinfo  {publisher} {Morgan Kaufmann Publishers Inc.},\ \bibinfo {address} {San Francisco, CA, USA},\ \bibinfo {year} {1986})\ pp.\ \bibinfo {pages} {128--137}\BibitemShut {NoStop}%
\bibitem [{\citenamefont {Tarjan}(1971)}]{dfs}%
  \BibitemOpen
  \bibfield  {author} {\bibinfo {author} {\bibfnamefont {R.}~\bibnamefont {Tarjan}},\ }\bibfield  {title} {\bibinfo {title} {Depth-first search and linear graph algorithms},\ }in\ \href {https://doi.org/10.1109/SWAT.1971.10} {\emph {\bibinfo {booktitle} {12th Annual Symposium on Switching and Automata Theory (swat 1971)}}}\ (\bibinfo {year} {1971})\ pp.\ \bibinfo {pages} {114--121}\BibitemShut {NoStop}%
\bibitem [{\citenamefont {Groff}\ and\ \citenamefont {Weinberg}(2009)}]{sql}%
  \BibitemOpen
  \bibfield  {author} {\bibinfo {author} {\bibfnamefont {J.}~\bibnamefont {Groff}}\ and\ \bibinfo {author} {\bibfnamefont {P.}~\bibnamefont {Weinberg}},\ }\href@noop {} {\emph {\bibinfo {title} {{SQL} {T}he {C}omplete {R}eference}}},\ \bibinfo {edition} {3rd}\ ed.\ (\bibinfo  {publisher} {McGraw-Hill, Inc.},\ \bibinfo {address} {USA},\ \bibinfo {year} {2009})\BibitemShut {NoStop}%
\bibitem [{\citenamefont {Codd}(1970)}]{codd1970relational}%
  \BibitemOpen
  \bibfield  {author} {\bibinfo {author} {\bibfnamefont {E.~F.}\ \bibnamefont {Codd}},\ }\bibfield  {title} {\bibinfo {title} {A relational model of data for large shared data banks},\ }\href@noop {} {\bibfield  {journal} {\bibinfo  {journal} {Communications of the ACM}\ }\textbf {\bibinfo {volume} {13}},\ \bibinfo {pages} {377} (\bibinfo {year} {1970})}\BibitemShut {NoStop}%
\bibitem [{\citenamefont {Schlag}\ \emph {et~al.}(2016)\citenamefont {Schlag}, \citenamefont {Henne}, \citenamefont {Heuer}, \citenamefont {Meyerhenke}, \citenamefont {Sanders},\ and\ \citenamefont {Schulz}}]{kahypar}%
  \BibitemOpen
  \bibfield  {author} {\bibinfo {author} {\bibfnamefont {S.}~\bibnamefont {Schlag}}, \bibinfo {author} {\bibfnamefont {V.}~\bibnamefont {Henne}}, \bibinfo {author} {\bibfnamefont {T.}~\bibnamefont {Heuer}}, \bibinfo {author} {\bibfnamefont {H.}~\bibnamefont {Meyerhenke}}, \bibinfo {author} {\bibfnamefont {P.}~\bibnamefont {Sanders}},\ and\ \bibinfo {author} {\bibfnamefont {C.}~\bibnamefont {Schulz}},\ }\bibfield  {title} {\bibinfo {title} {K-way hypergraph partitioning via n-level recursive bisection},\ }in\ \href {https://doi.org/10.1137/1.9781611974317.5} {\emph {\bibinfo {booktitle} {18th Workshop on Algorithm Engineering and Experiments, (ALENEX 2016)}}}\ (\bibinfo {year} {2016})\ pp.\ \bibinfo {pages} {53--67}\BibitemShut {NoStop}%
\bibitem [{\citenamefont {Arute}\ \emph {et~al.}(2019)\citenamefont {Arute}, \citenamefont {Arya}, \citenamefont {Babbush}, \citenamefont {Bacon}, \citenamefont {Bardin}, \citenamefont {Barends}, \citenamefont {Biswas}, \citenamefont {Boixo}, \citenamefont {Brandao}, \citenamefont {Buell}, \citenamefont {Burkett}, \citenamefont {Chen}, \citenamefont {Chen}, \citenamefont {Chiaro}, \citenamefont {Collins}, \citenamefont {Courtney}, \citenamefont {Dunsworth}, \citenamefont {Farhi}, \citenamefont {Foxen}, \citenamefont {Fowler}, \citenamefont {Gidney}, \citenamefont {Giustina}, \citenamefont {Graff}, \citenamefont {Guerin}, \citenamefont {Habegger}, \citenamefont {Harrigan}, \citenamefont {Hartmann}, \citenamefont {Ho}, \citenamefont {Hoffmann}, \citenamefont {Huang}, \citenamefont {Humble}, \citenamefont {Isakov}, \citenamefont {Jeffrey}, \citenamefont {Jiang}, \citenamefont {Kafri}, \citenamefont {Kechedzhi}, \citenamefont {Kelly}, \citenamefont {Klimov}, \citenamefont {Knysh}, \citenamefont {Korotkov},
  \citenamefont {Kostritsa}, \citenamefont {Landhuis}, \citenamefont {Lindmark}, \citenamefont {Lucero}, \citenamefont {Lyakh}, \citenamefont {Mandr{\`a}}, \citenamefont {McClean}, \citenamefont {McEwen}, \citenamefont {Megrant}, \citenamefont {Mi}, \citenamefont {Michielsen}, \citenamefont {Mohseni}, \citenamefont {Mutus}, \citenamefont {Naaman}, \citenamefont {Neeley}, \citenamefont {Neill}, \citenamefont {Niu}, \citenamefont {Ostby}, \citenamefont {Petukhov}, \citenamefont {Platt}, \citenamefont {Quintana}, \citenamefont {Rieffel}, \citenamefont {Roushan}, \citenamefont {Rubin}, \citenamefont {Sank}, \citenamefont {Satzinger}, \citenamefont {Smelyanskiy}, \citenamefont {Sung}, \citenamefont {Trevithick}, \citenamefont {Vainsencher}, \citenamefont {Villalonga}, \citenamefont {White}, \citenamefont {Yao}, \citenamefont {Yeh}, \citenamefont {Zalcman}, \citenamefont {Neven},\ and\ \citenamefont {Martinis}}]{sycamore}%
  \BibitemOpen
  \bibfield  {author} {\bibinfo {author} {\bibfnamefont {F.}~\bibnamefont {Arute}}, \bibinfo {author} {\bibfnamefont {K.}~\bibnamefont {Arya}}, \bibinfo {author} {\bibfnamefont {R.}~\bibnamefont {Babbush}}, \bibinfo {author} {\bibfnamefont {D.}~\bibnamefont {Bacon}}, \bibinfo {author} {\bibfnamefont {J.~C.}\ \bibnamefont {Bardin}}, \bibinfo {author} {\bibfnamefont {R.}~\bibnamefont {Barends}}, \bibinfo {author} {\bibfnamefont {R.}~\bibnamefont {Biswas}}, \bibinfo {author} {\bibfnamefont {S.}~\bibnamefont {Boixo}}, \bibinfo {author} {\bibfnamefont {F.~G. S.~L.}\ \bibnamefont {Brandao}}, \bibinfo {author} {\bibfnamefont {D.~A.}\ \bibnamefont {Buell}}, \bibinfo {author} {\bibfnamefont {B.}~\bibnamefont {Burkett}}, \bibinfo {author} {\bibfnamefont {Y.}~\bibnamefont {Chen}}, \bibinfo {author} {\bibfnamefont {Z.}~\bibnamefont {Chen}}, \bibinfo {author} {\bibfnamefont {B.}~\bibnamefont {Chiaro}}, \bibinfo {author} {\bibfnamefont {R.}~\bibnamefont {Collins}}, \bibinfo {author} {\bibfnamefont {W.}~\bibnamefont
  {Courtney}}, \bibinfo {author} {\bibfnamefont {A.}~\bibnamefont {Dunsworth}}, \bibinfo {author} {\bibfnamefont {E.}~\bibnamefont {Farhi}}, \bibinfo {author} {\bibfnamefont {B.}~\bibnamefont {Foxen}}, \bibinfo {author} {\bibfnamefont {A.}~\bibnamefont {Fowler}}, \bibinfo {author} {\bibfnamefont {C.}~\bibnamefont {Gidney}}, \bibinfo {author} {\bibfnamefont {M.}~\bibnamefont {Giustina}}, \bibinfo {author} {\bibfnamefont {R.}~\bibnamefont {Graff}}, \bibinfo {author} {\bibfnamefont {K.}~\bibnamefont {Guerin}}, \bibinfo {author} {\bibfnamefont {S.}~\bibnamefont {Habegger}}, \bibinfo {author} {\bibfnamefont {M.~P.}\ \bibnamefont {Harrigan}}, \bibinfo {author} {\bibfnamefont {M.~J.}\ \bibnamefont {Hartmann}}, \bibinfo {author} {\bibfnamefont {A.}~\bibnamefont {Ho}}, \bibinfo {author} {\bibfnamefont {M.}~\bibnamefont {Hoffmann}}, \bibinfo {author} {\bibfnamefont {T.}~\bibnamefont {Huang}}, \bibinfo {author} {\bibfnamefont {T.~S.}\ \bibnamefont {Humble}}, \bibinfo {author} {\bibfnamefont {S.~V.}\ \bibnamefont
  {Isakov}}, \bibinfo {author} {\bibfnamefont {E.}~\bibnamefont {Jeffrey}}, \bibinfo {author} {\bibfnamefont {Z.}~\bibnamefont {Jiang}}, \bibinfo {author} {\bibfnamefont {D.}~\bibnamefont {Kafri}}, \bibinfo {author} {\bibfnamefont {K.}~\bibnamefont {Kechedzhi}}, \bibinfo {author} {\bibfnamefont {J.}~\bibnamefont {Kelly}}, \bibinfo {author} {\bibfnamefont {P.~V.}\ \bibnamefont {Klimov}}, \bibinfo {author} {\bibfnamefont {S.}~\bibnamefont {Knysh}}, \bibinfo {author} {\bibfnamefont {A.}~\bibnamefont {Korotkov}}, \bibinfo {author} {\bibfnamefont {F.}~\bibnamefont {Kostritsa}}, \bibinfo {author} {\bibfnamefont {D.}~\bibnamefont {Landhuis}}, \bibinfo {author} {\bibfnamefont {M.}~\bibnamefont {Lindmark}}, \bibinfo {author} {\bibfnamefont {E.}~\bibnamefont {Lucero}}, \bibinfo {author} {\bibfnamefont {D.}~\bibnamefont {Lyakh}}, \bibinfo {author} {\bibfnamefont {S.}~\bibnamefont {Mandr{\`a}}}, \bibinfo {author} {\bibfnamefont {J.~R.}\ \bibnamefont {McClean}}, \bibinfo {author} {\bibfnamefont {M.}~\bibnamefont
  {McEwen}}, \bibinfo {author} {\bibfnamefont {A.}~\bibnamefont {Megrant}}, \bibinfo {author} {\bibfnamefont {X.}~\bibnamefont {Mi}}, \bibinfo {author} {\bibfnamefont {K.}~\bibnamefont {Michielsen}}, \bibinfo {author} {\bibfnamefont {M.}~\bibnamefont {Mohseni}}, \bibinfo {author} {\bibfnamefont {J.}~\bibnamefont {Mutus}}, \bibinfo {author} {\bibfnamefont {O.}~\bibnamefont {Naaman}}, \bibinfo {author} {\bibfnamefont {M.}~\bibnamefont {Neeley}}, \bibinfo {author} {\bibfnamefont {C.}~\bibnamefont {Neill}}, \bibinfo {author} {\bibfnamefont {M.~Y.}\ \bibnamefont {Niu}}, \bibinfo {author} {\bibfnamefont {E.}~\bibnamefont {Ostby}}, \bibinfo {author} {\bibfnamefont {A.}~\bibnamefont {Petukhov}}, \bibinfo {author} {\bibfnamefont {J.~C.}\ \bibnamefont {Platt}}, \bibinfo {author} {\bibfnamefont {C.}~\bibnamefont {Quintana}}, \bibinfo {author} {\bibfnamefont {E.~G.}\ \bibnamefont {Rieffel}}, \bibinfo {author} {\bibfnamefont {P.}~\bibnamefont {Roushan}}, \bibinfo {author} {\bibfnamefont {N.~C.}\ \bibnamefont {Rubin}},
  \bibinfo {author} {\bibfnamefont {D.}~\bibnamefont {Sank}}, \bibinfo {author} {\bibfnamefont {K.~J.}\ \bibnamefont {Satzinger}}, \bibinfo {author} {\bibfnamefont {V.}~\bibnamefont {Smelyanskiy}}, \bibinfo {author} {\bibfnamefont {K.~J.}\ \bibnamefont {Sung}}, \bibinfo {author} {\bibfnamefont {M.~D.}\ \bibnamefont {Trevithick}}, \bibinfo {author} {\bibfnamefont {A.}~\bibnamefont {Vainsencher}}, \bibinfo {author} {\bibfnamefont {B.}~\bibnamefont {Villalonga}}, \bibinfo {author} {\bibfnamefont {T.}~\bibnamefont {White}}, \bibinfo {author} {\bibfnamefont {Z.~J.}\ \bibnamefont {Yao}}, \bibinfo {author} {\bibfnamefont {P.}~\bibnamefont {Yeh}}, \bibinfo {author} {\bibfnamefont {A.}~\bibnamefont {Zalcman}}, \bibinfo {author} {\bibfnamefont {H.}~\bibnamefont {Neven}},\ and\ \bibinfo {author} {\bibfnamefont {J.~M.}\ \bibnamefont {Martinis}},\ }\bibfield  {title} {\bibinfo {title} {Quantum supremacy using a programmable superconducting processor},\ }\href {https://doi.org/10.1038/s41586-019-1666-5} {\bibfield
  {journal} {\bibinfo  {journal} {Nature}\ }\textbf {\bibinfo {volume} {574}},\ \bibinfo {pages} {505} (\bibinfo {year} {2019})}\BibitemShut {NoStop}%
\bibitem [{\citenamefont {Bauernfeind}\ \emph {et~al.}(2017)\citenamefont {Bauernfeind}, \citenamefont {Zingl}, \citenamefont {Triebl}, \citenamefont {Aichhorn},\ and\ \citenamefont {Evertz}}]{ftps}%
  \BibitemOpen
  \bibfield  {author} {\bibinfo {author} {\bibfnamefont {D.}~\bibnamefont {Bauernfeind}}, \bibinfo {author} {\bibfnamefont {M.}~\bibnamefont {Zingl}}, \bibinfo {author} {\bibfnamefont {R.}~\bibnamefont {Triebl}}, \bibinfo {author} {\bibfnamefont {M.}~\bibnamefont {Aichhorn}},\ and\ \bibinfo {author} {\bibfnamefont {H.~G.}\ \bibnamefont {Evertz}},\ }\bibfield  {title} {\bibinfo {title} {Fork tensor-product states: efficient multiorbital real-time {DMFT} solver},\ }\bibfield  {journal} {\bibinfo  {journal} {Phys. Rev. X}\ }\textbf {\bibinfo {volume} {7}},\ \href {https://doi.org/10.1103/physrevx.7.031013} {10.1103/physrevx.7.031013} (\bibinfo {year} {2017})\BibitemShut {NoStop}%
\bibitem [{\citenamefont {Smith}\ and\ \citenamefont {Gray}(2018)}]{opt_einsum}%
  \BibitemOpen
  \bibfield  {author} {\bibinfo {author} {\bibfnamefont {D.~G.~A.}\ \bibnamefont {Smith}}\ and\ \bibinfo {author} {\bibfnamefont {J.}~\bibnamefont {Gray}},\ }\bibfield  {title} {\bibinfo {title} {opt\_einsum - a {P}ython package for optimizing contraction order for einsum-like expressions},\ }\href {https://doi.org/10.21105/joss.00753} {\bibfield  {journal} {\bibinfo  {journal} {Journal of Open Source Software}\ }\textbf {\bibinfo {volume} {3}},\ \bibinfo {pages} {753} (\bibinfo {year} {2018})}\BibitemShut {NoStop}%
\bibitem [{\citenamefont {Boixo}\ \emph {et~al.}(2018{\natexlab{b}})\citenamefont {Boixo}, \citenamefont {Isakov}, \citenamefont {Smelyanskiy}, \citenamefont {Babbush}, \citenamefont {Ding}, \citenamefont {Jiang}, \citenamefont {Bremner}, \citenamefont {Martinis},\ and\ \citenamefont {Neven}}]{bristlecone}%
  \BibitemOpen
  \bibfield  {author} {\bibinfo {author} {\bibfnamefont {S.}~\bibnamefont {Boixo}}, \bibinfo {author} {\bibfnamefont {S.~V.}\ \bibnamefont {Isakov}}, \bibinfo {author} {\bibfnamefont {V.~N.}\ \bibnamefont {Smelyanskiy}}, \bibinfo {author} {\bibfnamefont {R.}~\bibnamefont {Babbush}}, \bibinfo {author} {\bibfnamefont {N.}~\bibnamefont {Ding}}, \bibinfo {author} {\bibfnamefont {Z.}~\bibnamefont {Jiang}}, \bibinfo {author} {\bibfnamefont {M.~J.}\ \bibnamefont {Bremner}}, \bibinfo {author} {\bibfnamefont {J.~M.}\ \bibnamefont {Martinis}},\ and\ \bibinfo {author} {\bibfnamefont {H.}~\bibnamefont {Neven}},\ }\bibfield  {title} {\bibinfo {title} {Characterizing quantum supremacy in near-term devices},\ }\href {https://doi.org/10.1038/s41567-018-0124-x} {\bibfield  {journal} {\bibinfo  {journal} {Nat. Phys.}\ }\textbf {\bibinfo {volume} {14}},\ \bibinfo {pages} {595} (\bibinfo {year} {2018}{\natexlab{b}})}\BibitemShut {NoStop}%
\bibitem [{\citenamefont {Radke}\ and\ \citenamefont {Neumann}(2019)}]{radke}%
  \BibitemOpen
  \bibfield  {author} {\bibinfo {author} {\bibfnamefont {B.}~\bibnamefont {Radke}}\ and\ \bibinfo {author} {\bibfnamefont {T.}~\bibnamefont {Neumann}},\ }\bibfield  {title} {\bibinfo {title} {Lindp++: Generalizing linearized {DP} to crossproducts and non-inner joins},\ }in\ \href {https://doi.org/10.18420/btw2019-05} {\emph {\bibinfo {booktitle} {Datenbanksysteme f{\"{u}}r Business, Technologie und Web {(BTW} 2019), 18. Fachtagung des GI-Fachbereichs ,,Datenbanken und Informationssysteme" (DBIS), 4.-8. M{\"{a}}rz 2019, Rostock, Germany, Proceedings}}},\ \bibinfo {series} {LNI}, Vol.\ \bibinfo {volume} {P-289},\ \bibinfo {editor} {edited by\ \bibinfo {editor} {\bibfnamefont {T.}~\bibnamefont {Grust}}, \bibinfo {editor} {\bibfnamefont {F.}~\bibnamefont {Naumann}}, \bibinfo {editor} {\bibfnamefont {A.}~\bibnamefont {B{\"{o}}hm}}, \bibinfo {editor} {\bibfnamefont {W.}~\bibnamefont {Lehner}}, \bibinfo {editor} {\bibfnamefont {T.}~\bibnamefont {H{\"{a}}rder}}, \bibinfo {editor} {\bibfnamefont {E.}~\bibnamefont
  {Rahm}}, \bibinfo {editor} {\bibfnamefont {A.}~\bibnamefont {Heuer}}, \bibinfo {editor} {\bibfnamefont {M.}~\bibnamefont {Klettke}},\ and\ \bibinfo {editor} {\bibfnamefont {H.}~\bibnamefont {Meyer}}}\ (\bibinfo  {publisher} {Gesellschaft f{\"{u}}r Informatik, Bonn},\ \bibinfo {year} {2019})\ pp.\ \bibinfo {pages} {57--76}\BibitemShut {NoStop}%
\bibitem [{\citenamefont {Huang}\ \emph {et~al.}(2018)\citenamefont {Huang}, \citenamefont {Matthews},\ and\ \citenamefont {van~de Geijn}}]{Huang2018}%
  \BibitemOpen
  \bibfield  {author} {\bibinfo {author} {\bibfnamefont {J.}~\bibnamefont {Huang}}, \bibinfo {author} {\bibfnamefont {D.~A.}\ \bibnamefont {Matthews}},\ and\ \bibinfo {author} {\bibfnamefont {R.~A.}\ \bibnamefont {van~de Geijn}},\ }\bibfield  {title} {\bibinfo {title} {Strassen's algorithm for tensor contraction},\ }\href {https://doi.org/10.1137/17M1135578} {\bibfield  {journal} {\bibinfo  {journal} {SIAM J. Sci. Comput.}\ }\textbf {\bibinfo {volume} {40}},\ \bibinfo {pages} {C305} (\bibinfo {year} {2018})}\BibitemShut {NoStop}%
\bibitem [{\citenamefont {Bellman}(1952)}]{bellman}%
  \BibitemOpen
  \bibfield  {author} {\bibinfo {author} {\bibfnamefont {R.}~\bibnamefont {Bellman}},\ }\bibfield  {title} {\bibinfo {title} {On the theory of dynamic programming},\ }\href {http://www.jstor.org/stable/88493} {\bibfield  {journal} {\bibinfo  {journal} {Proc. Natl. Acad. Sci. USA}\ }\textbf {\bibinfo {volume} {38}},\ \bibinfo {pages} {716} (\bibinfo {year} {1952})}\BibitemShut {NoStop}%
\end{thebibliography}%


\clearpage
\appendix
\onecolumngrid


\section{Proofs}


\subsection{Theorem~\ref{theorem:asi_cost_function}}\label{appendix:proof_theorem_1}

\begin{proof}

We recursively expand $\C(AUVB)$ and $\C(AVUB)$ using Eq.~\eqref{eq:other_cout}:

\begin{align*}
&\qquad&\C(AUVB) &\leq \C(AVUB) &&\\ 
\overset{\eqref{eq:other_cout}}{\iff}&& \C(AUV) + \frac{\overline{\#}{T^{[AUV]}}}{|e_{AUV,B}|}\C(B) &\leq \C(AVU) + \frac{\overline{\#}{T^{[{AVU}]}}}{|e_{AVU,B}|}\C(B) && 
\intertext{As both $T^{[AUV]}$ and $T^{[AVU]}$ represent the same tensor and both $e_{AUV,B}$ and $e_{AVU,B}$ refer to the same leg, we have}
\iff&&\C(AUV) &\leq \C(AVU) &&\\ 
\overset{\eqref{eq:other_cout}}{\iff}&& \C(AU) + \frac{\overline{\#}{T^{[AU]}}}{|e_{AU,V}|}\C(V) &\leq \C(AV) + \frac{\overline{\#}{T^{[AV]}}}{|e_{AV,U}|}\C(U) &&\\ 
\overset{\eqref{eq:other_cout}}{\iff}&& \C(A) + \frac{\overline{\#}{T^{[A]}}}{|e_{A,U}|}\C(U) + \frac{\overline{\#}{T^{[AU]}}}{|e_{AU,V}|}\C(V) &\leq \C(A) + \frac{\overline{\#}{T^{[A]}}}{|e_{A,V}|}\C(V) + \frac{\overline{\#}{T^{[AV]}}}{|e_{AV,U}|}\C(U) &&\\ 
\iff&& \frac{\overline{\#}{T^{[A]}}}{|e_{A,U}|}\C(U) + \frac{\overline{\#}{T^{[AU]}}}{|e_{AU,V}|}\C(V) &\leq \frac{\overline{\#}{T^{[A]}}}{|e_{A,V}|}\C(V) + \frac{\overline{\#}{T^{[AV]}}}{|e_{AV,U}|}\C(U)&&
\intertext{Note that $\overline{\#}{T^{[AU]}} = \frac{\overline{\#}{T^{[A]}}\overline{\#}{T^{[U]}}}{|e_{A,U}|}$, due to Eq.~\eqref{eq:norm_identity}. Analogous for $\overline{\#}{T^{[AV]}}$. Hence, we obtain}
\iff&& \frac{\overline{\#}{T^{[A]}}}{|e_{A,U}|}\C(U) + \frac{\cg{trueblue}{\overline{\#}{T^{[A]}}\overline{\#}{T^{[U]}}}}{|e_{AU,V}|\cg{trueblue}{|e_{A,U}|}}\C(V) &\leq \frac{\overline{\#}{T^{[A]}}}{|e_{A,V}|}\C(V) + \frac{\cg{trueblue}{\overline{\#}{T^{[A]}}\overline{\#}{T^{[V]}}}}{|e_{AV,U}|\cg{trueblue}{|e_{A,V}|}}\C(U) &&\\ 
\iff&& \frac{\C(U)}{|e_{A,U}|} + \frac{\overline{\#}{T^{[U]}}}{|e_{AU,V}||e_{A,U}|}\C(V) &\leq \frac{\C(V)}{|e_{A,V}|} + \frac{\overline{\#}{T^{[V]}}}{|e_{AV,U}||e_{A,V}|}\C(U) && \intertext{Next, since $AUV$ and $AVU$ must both satisfy the constraints of the precedence graph due to Def.~\ref{def:asi}, there is no edge between $U$ and $V$. This is because the underlying precedence graph is a rooted tree, so we cannot have a directed edge between $U$ and $V$ \emph{and} between $V$ and $U$ at the same time, otherwise, a cycle would form. This implies that the edge $e_{AU,V}$ is simply the edge $e_{A,V}$. Analogously for $e_{AV,U}$.}
\iff&& \frac{\C(U)}{|e_{A,U}|} + \frac{\overline{\#}{T^{[U]}}}{|e_{\cg{persimmon}{A,V}}||e_{A,U}|}\C(V) &\leq \frac{\C(V)}{|e_{A,V}|} + \frac{\overline{\#}{T^{[V]}}}{|e_{\cg{persimmon}{A,U}}||e_{A,V}|}\C(U) &&\\ 
\iff&& \frac{\C(U)}{|e_{A,U}|}\left(1 - \frac{\overline{\#}{T^{[V]}}}{|e_{A,V}|}\right) &\leq \frac{\C(V)}{|e_{A,V}|}\left(1 - \frac{\overline{\#}{T^{[U]}}}{|e_{A,U}|}\right) &&\\ 
\iff&& \C(U)\left(|e_{A,V}| - \overline{\#}{T^{[V]}}\right) &\leq \C(V)\left(|e_{A,U}| - \overline{\#}{T^{[U]}}\right).
\end{align*}
As mentioned in Sec.~\ref{subsec:simplifications}, the edges $e_{A,U}$ and $e_{A,V}$ are the unique tree edges to $U$ and $V$, and we can denote them instead by $e_{U}$ and $e_{V}$, respectively. Hence, the inequality reduces to one where only functions of $U$ and $V$ are involved. Defining $\sigma(S)$ as $(\C(S), |e_S| - \overline{\#}T^{[S]})$ completes the proof.

\end{proof}


\subsection{Theorem~\ref{theorem:normalize}}\label{appendix:proof_theorem_2}

\begin{proof}
Since $A \rightarrow B$, any optimal sequence must be of the form $UAVBW$, with $U \neq \varnothing$ (empty sequence).~\footnote{This condition is necessary due to the definition of $\sigma$. Namely, $\sigma$, used in the sequel to evaluate the $\sigma(A)$, is not defined on sequences starting with the root of the precedence graph. This relates to the fact that it is not necessary to normalize the linearization of the root of the precedence graph, as explained in Sec.~\ref{subsec:simplifications}.} In case $V = \varnothing$, there is nothing more to prove, otherwise we distinguish between the following subcases:
\begin{description}[before={\renewcommand\makelabel[1]{\bfseries ##1}}]
\item[Case] $A \rightarrow V$:
\begin{description}[before={\renewcommand\makelabel[1]{\bfseries ##1}}]
\item[Case] $\sigma(A) \geq \sigma(V)$: This is not possible, otherwise $V$ would have been previously absorbed in a compound tensor with $A$.
\item[Case] $\sigma(A) < \sigma(V)$: Since $\sigma(B) \leq \sigma(A)$, by transitivity we have $\sigma(B) < \sigma(V)$. Note that $V \rightarrow B$ is not possible, otherwise $V$ would have been previously absorbed in a compound tensor with $B$. Hence $V \not\rightarrow B$ and we can exchange $B$ and $V$ without increasing the cost. 
\end{description}
\item[Case] $A \not\rightarrow V$:
\begin{description}[before={\renewcommand\makelabel[1]{\bfseries ##1}}]
\item[Case] $\sigma(A) \geq \sigma(V)$: We can exchange $V$ and $A$ without increasing the costs.
\item[Case] $\sigma(A) < \sigma(V)$: Since $\sigma(B) \leq \sigma(A)$, by transitivity we have $\sigma(B) < \sigma(V)$. Since $V \not\rightarrow B$, otherwise the tree constraint would be violated, we can exchange $B$ and $V$ without increasing the cost. 
\end{description}
\end{description}
\end{proof}


\subsection{Transitivity of ``$\leq$''}\label{appendix:proof_transitivity}

\begin{proof}
Let $f_1 = (a, b)$, $f_2 = (c, d)$, and $f_3 = (e, f)$ be three elements in $\mathbb{N}^+ \times \mathbb{Z}$. We have to prove
\begin{equation*}
    (a, b) \leq (c, d) \land (c, d) \leq (e, f) \implies (a, b) \leq (e, f).
\end{equation*}
By the definition of ``$\leq$'', we are left with proving
\begin{equation*}
    ad \leq bc \land cf \leq de \implies af \leq be,
\end{equation*}
which follows from
\begin{equation*}
cf \leq de \overset{c>0}{\iff} f \leq \frac{de}{c} \overset{a > 0}{\iff} af \leq \frac{ade}{c} \overset{ad\leq bc}{\implies} af \leq \frac{bce}{c} \iff af \leq be.
\end{equation*}
\end{proof}


\section{{\normalsize \texttt{LinDP}}: Extended Pseudocode}\label{appendix:lindp}

In Sec.~\ref{subsec:lindp} we only provided a shortened version of the pseudocode, as the underlying dynamic program is exactly the solution to the well-known matrix-chain multiplication problem~\cite{cormen}. In Alg.~\ref{algo:lindp_extended}, its extended version is outlined.

\begin{algorithm}
  \caption{\texttt{LinDP}~\cite{adaptive} adapted for tensor networks (extended version)}\label{algo:lindp_extended}
  \begin{algorithmic}[1]
  \STATE \textbf{Input:} Tensor network $\mathcal{T} = (V, E, c)$
  \STATE \textbf{Output:} General contraction tree $\tau$
  \STATE $\tau^* \leftarrow \boldsymbol\varnothing$
  \FOR{\textbf{each} $T^{[i]}$ in $V$}
      \STATE $S \leftarrow$ Linearization of precedence graph $\mathcal{P}_{T^{[i]}}$ cf. \texttt{TensorIKKBZ} (Alg.~\ref{algo:tensor_ikkbz})
      \STATE $\texttt{dp}[i,\:i] = 0, \forall i\in[n]$
      \FOR {\textbf{each} $s \in [2, \ldots, n]$}
            \FOR {\textbf{each} $i \in [1, \ldots, n - s + 1]$}
                \STATE $j \leftarrow i + s - 1$
                \FOR {\textbf{each} $k \in [i, j[$}
                    \STATE $c' \leftarrow \texttt{dp}[i,\:k] + \texttt{dp}[k + 1,\:j] + c(T^{[S_i, \ldots, S_k]}, T^{[S_{k + 1}, \ldots, S_j]})$
                    \IF {$c' < \texttt{dp}[i, \:j]$}
                        \STATE $\texttt{dp}[i, \:j] = c'$
                        \STATE $\texttt{opt}[i, \:j] = k$
                    \ENDIF
                \ENDFOR
            \ENDFOR
      \ENDFOR
      \IF {$\tau^* = \varnothing$ or $\texttt{dp}[1, \:n] < \C(\tau^*)$}
            \STATE $\tau^* \leftarrow$ Reconstruct contraction tree from $\texttt{opt}$
      \ENDIF
  \ENDFOR
  \RETURN $\tau^*$
  \end{algorithmic}
\end{algorithm}

The algorithm aims to find the optimal contraction tree respecting an initial linear order (sequence) $S$ of the tensors. To this end, it exploits Bellman's optimality condition~\cite{bellman}, which states that if a problem $\mathcal{P}$ has an optimal solution $\mathcal{S}^* := \mathcal{S}_1\mathcal{S}_2$, then both $\mathcal{S}_1$ and $\mathcal{S}_2$ are optimal solutions for $\mathcal{P}_1$ and $\mathcal{P}_2$, respectively, with $\mathcal{P} = \mathcal{P}_1\mathcal{P}_2$. In our context, this means that every contraction subtree of the original contraction tree must be an optimal contraction tree for the tensors it contains.

Once the precedence graph $\mathcal{P}_{T^{[i]}}$ of $T^{[i]}$ has been linearized, the key idea is to compute $\C(\cdot)$ for all intervals $[i,\:j],\:i \leq j$, using Eq.~\eqref{eq:contr_cost_rec}. These values are stored in the dynamic programming table $\texttt{dp}$ of size $n \times n$. For the base case in Eq.~\eqref{eq:contr_cost_rec}, we initialize $\texttt{dp}[i, i]$ to 0, $\forall i\in[n]$ (line 6). Subsequently, we fix an interval size $s$ and iterate all intervals of that size. Then, by employing Bellman's optimality principle, we have to guess where the problem $[i, j]$ splits. This is represented by the index $k$, for each of which we compute the cost (cf. Eq.~\eqref{eq:contr_cost_rec}) and compare with the current minimum obtained for $[i, j]$ (lines 11-12). If the current $k$ improves the cost, we store it in the table $\texttt{opt}$ (of size $n\times n$) which will be used to reconstruct the contraction tree of optimal cost. Finally, after the cubic-time algorithm has finished, we update the contraction tree $\tau$ in case the cost $\texttt{dp}[1,\:n]$, the optimal contraction cost for the given linearization, is better. If so, we reconstruct the contraction tree using the table $\texttt{opt}$ (line 20).

\section{Query Graph}\label{appendix:query_graph}

\begin{figure}
    \centering
    \begin{minipage}[b]{0.45\textwidth}
        \texttt{\small{SELECT s.id\\\qquad FROM students s, exams e, professors p\\\qquad WHERE \textcolor{persimmon}{s.id = e.student\_id}\\\qquad\quad AND \textcolor{darkblue}{p.id = e.prof\_id}\\\qquad\quad AND p.name = 'R. Feynman'}}
    \end{minipage}
    \raisebox{-2\baselineskip}{
        \centering
        \includegraphics[]{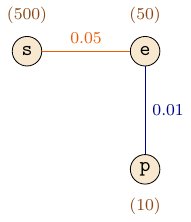}
    }
\caption{Example SQL query and its corresponding query graph. The vertices correspond to the tables in the query, while the edges correspond to the joins between them.}
\label{fig:sql_query_graph}
\end{figure}

A query graph is a visual representation of a query. The tables are represented by the vertices of the graph, while a join between two tables is represented by an edge between the corresponding vertices. To illustrate this construction, consider the SQL query in Fig.~\ref{fig:sql_query_graph}. Its output is the list of ids of all students who have taken an exam with R.~Feynman. This is done by two joins: one between the \texttt{students} and \texttt{exams}, and one between \texttt{exams} and \texttt{professors}. The vertices of the query graph are thus the three tables \texttt{students}, \texttt{exams}, and \texttt{professors}, respectively, while the two joins are represented as edges between them. The vertex labels correspond to the cardinalities of the tables, while the edge weights correspond to the join selectivities, which we formally introduced in Sec.~\ref{subsec:commonalities}.

Join selectivities are used to compute the cardinality of sets of relations. Formally, the cardinality of a set $S$ of relations is the product of individual relation cardinalities and the join selectivities associated with the edges between them. For instance, the cardinality of $s \Join e$ is equal to $500 \cdot 50 \cdot 0.05 = 1250$, while the cardinality of the entire query $s \Join e \Join p$ is $500 \cdot 50 \cdot 10 \cdot 0.05 \cdot 0.01 = 125$. In Sec.~\ref{subsec:commonalities}, we also mentioned the term \emph{cross product}. A cross product can be thought of as a join between two tables that do not share a join edge. For example, there is no join edge between $s$ and $p$, resulting in a cross product; its cardinality is therefore $500 \cdot 10 = 5000$.


\section{Examples}


\subsection{Tensor Network Contraction Cost}\label{appendix:contraction_cost}

We exemplify the calculation of $\mathcal{C}$ for the linear contraction order in Fig.~\ref{fig:example_linear_contraction}. For simplicity, assume all legs are of size 2. In the first step, the cost is $2 \cdot 2 \cdot 2 = 8$, as there are three legs involved in the operation. Next, the contraction involves all legs, hence the cost equals again 8. Lastly, the cost is $2 \cdot 2$, as there are two legs between the remaining two tensors. Therefore, the contraction cost equals $8 + 8 + 4 = 20$.

\subsection{ASI Example: Full Derivation}\label{appendix:asi_derivation}

We provide the full derivation for the expressions used in Sec.~\ref{sec:asi_motivation}. Namely, we calculate the costs of $T^{[1]}\cg{trueblue}{T^{[2]}}\cg{persimmon}{T^{[4]}}T^{[3]}$ and $T^{[1]}\cg{persimmon}{T^{[4]}}\cg{trueblue}{T^{[2]}}T^{[3]}$ using Eq.~\eqref{eq:contr_cost_rec} to recursively decompose the cost until we arrive at contractions between individual tensors, as follows:
\begin{flalign*}
\mathcal{C}(T^{[1]}\cg{trueblue}{T^{[2]}}\cg{persimmon}{T^{[4]}}T^{[3]}) &= \mathcal{C}(T^{[1]}T^{[2]}T^{[4]}) + \mathcal{C}(T^{[3]}) + c(T^{[1,2,4]}, T^{[3]})\\
&= \mathcal{C}(T^{[1]}T^{[2]}T^{[4]}) + 0 + c(T^{[1,2,4]}, T^{[3]})\\
&= \mathcal{C}(T^{[1]}T^{[2]}) + \mathcal{C}(T^{[4]}) + c(T^{[1,2]}, T^{[4]}) + c(T^{[1,2,4]}, T^{[3]})\\
&= \mathcal{C}(T^{[1]}T^{[2]}) + 0 + c(T^{[1,2]}, T^{[4]}) + c(T^{[1,2,4]}, T^{[3]})\\
&= \mathcal{C}(T^{[1]}) + \mathcal{C}(T^{[2]}) + c(T^{[1]}, T^{[2]}) + c(T^{[1,2]}, T^{[4]}) + c(T^{[1,2,4]}, T^{[3]})\\
&= 0 + 0 + c(T^{[1]}, T^{[2]}) + c(T^{[1,2]}, T^{[4]}) + c(T^{[1,2,4]}, T^{[3]})\\
&= c(T^{[1]}, T^{[2]}) + c(T^{[1,2]}, T^{[4]}) + c(T^{[1,2,4]}, T^{[3]})\\
&= pqr + qr + r
\end{flalign*}
\begin{flalign*}
\mathcal{C}(T^{[1]}\cg{persimmon}{T^{[4]}}\cg{trueblue}{T^{[2]}}T^{[3]}) &= \mathcal{C}(T^{[1]}T^{[4]}T^{[2]}) + \mathcal{C}(T^{[3]}) + c(T^{[1,4,2]}, T^{[3]})\\
&= \mathcal{C}(T^{[1]}T^{[4]}T^{[2]}) + 0 + c(T^{[1,4,2]}, T^{[3]})\\
&= \mathcal{C}(T^{[1]}T^{[4]}) + \mathcal{C}(T^{[2]}) + c(T^{[1,4]}, T^{[2]}) + c(T^{[1,4,2]}, T^{[3]})\\
&= \mathcal{C}(T^{[1]}T^{[4]}) + 0 + c(T^{[1,4]}, T^{[2]}) + c(T^{[1,4,2]}, T^{[3]})\\
&= \mathcal{C}(T^{[1]}) + \mathcal{C}(T^{[4]}) + c(T^{[1]}, T^{[4]}) + c(T^{[1,4]}, T^{[2]}) + c(T^{[1,4,2]}, T^{[3]})\\
&= 0 + 0 + c(T^{[1]}, T^{[4]}) + c(T^{[1,4]}, T^{[2]}) + c(T^{[1,4,2]}, T^{[3]})\\
&= c(T^{[1]}, T^{[4]}) + c(T^{[1,4]}, T^{[2]}) + c(T^{[1,4,2]}, T^{[3]})\\
&= pq + pr + r.
\end{flalign*}


\subsection{Exemplification of $\C(S)$}\label{appendix:recursive_example}

\begin{figure}
    \centering
    \begin{subfigure}[c]{0.4\columnwidth}
        \centering
        \includegraphics[]{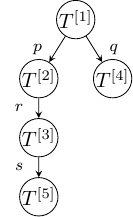}
        \caption{Rooted tree tensor network}
        \label{subfig:rooted_ttn}
    \end{subfigure}
    \begin{subfigure}[c]{0.4\columnwidth}
    \centering
        \begin{flalign*}
            \mathcal{C}(T^{[2]}T^{[3]}) &= \mathcal{C}(T^{[2]}) + \frac{\overline{\#}T^{[2]}}{|e_{T^{[3]}}|}\mathcal{C}(T^{[3]})\\
            &= \#T^{[2]} + \frac{\overline{\#}T^{[2]}}{|e_{T^{[3]}}|}\#T^{[3]}\\
            &= pr + \frac{r}{r} \cdot rs\\
            &= pr + rs.
        \end{flalign*}
        \caption{Computing $\C(T^{[2]}T^{[3]})$}
        \label{subfig:example_cost_function_small}
    \end{subfigure}
    \caption[]{Example for computing $\C$ cf. Eq.~\eqref{eq:other_cout}}
    \label{fig:cost_function_example}
\end{figure}
We exemplify the computation of $\C(S)$, as specified in the recursive Eq.~\eqref{eq:other_cout}, with the (rooted) tree tensor network depicted in Fig.~\ref{subfig:rooted_ttn}. Let us first compute the cost of contracting $T^{[2]}$ with $T^{[3]}$, and observe that Eq.~\eqref{eq:other_cout} corresponds to the actual contraction cost of two tensors. A step-wise computation is illustrated in Fig.~\ref{subfig:example_cost_function_small}. Let us now compute the contraction cost of contracting $T^{[1]}$ with $T^{[2]}$ and then with $T^{[3]}$ using the previously computed $\C(T^{[2]}T^{[3]})$:
\begin{flalign*}
    \mathcal{C}(T^{[1]}T^{[2]}T^{[3]}) &= \mathcal{C}(T^{[1]}) + \frac{\overline{\#}T^{[1]}}{|e_{T^{[2]}}|}\mathcal{C}(T^{[2]}T^{[3]})\\&= 0 + \frac{pq}{p}(pr + rs)\\&= qpr + qrs.
\end{flalign*}
\label{eq:example_2}
To underscore the fact that the splitting point of the sequence does not matter for the contraction cost, let us consider the split $T^{[1]}T^{[2]}T^{[3]} := (T^{[1]}T^{[2]})(T^{[3]})$ instead of the previous one, namely $T^{[1]}T^{[2]}T^{[3]} := (T^{[1]})(T^{[2]}T^{[3]})$:
\begin{flalign*}
    \mathcal{C}(T^{[1]}T^{[2]}T^{[3]}) =&\:\mathcal{C}(T^{[1]}T^{[2]}) + \frac{\overline{\#}T^{[1,2]}}{|e_{T^{[3]}}|}\mathcal{C}(T^{[3]})\\
    =&\:\C(T^{[1]}) + \frac{\overline{\#}T^{[1]}}{|e_{T^{[2]}}|}\C(T^{[2]}) + \frac{\overline{\#}T^{[1,2]}}{|e_{T^{[3]}}|}\#T^{[3]}\\
    \overset{\eqref{eq:norm_identity}}{=}&\:0 + \frac{\overline{\#}T^{[1]}}{|e_{T^{[2]}}|}\#T^{[2]} + \frac{\overline{\#}T^{[1]}\overline{\#}T^{[2]}}{|e_{T^{[3]}}||e_{T^{[2]}}|}\#T^{[3]}\\
    =&\:0 + \frac{pq}{p} \cdot pr + \frac{pq \cdot r}{r \cdot p} \cdot rs\\
    =&\:qpr + qrs.
\end{flalign*}


\subsection{\texttt{TensorIKKBZ}: Example}\label{appendix:tensor_ikkbz_example}

\begin{figure}
    \centering
    \includegraphics{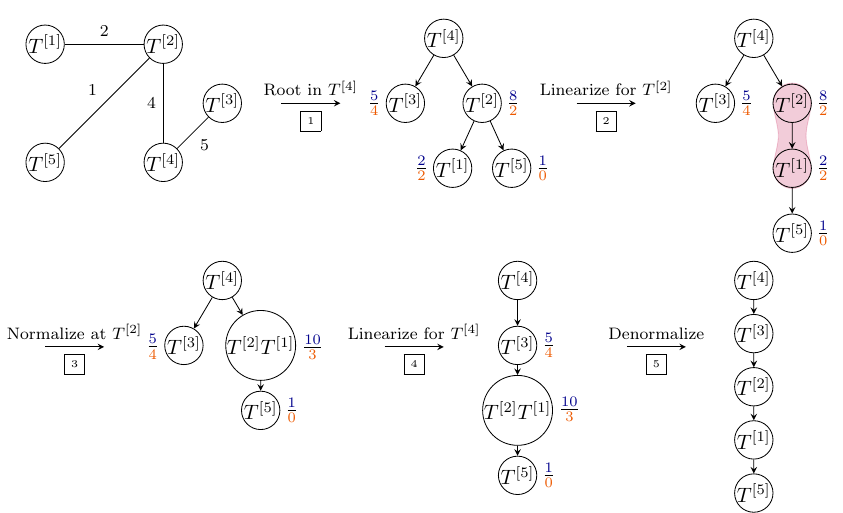}
    \caption[\texttt{TensorIKKBZ}: Example]{\texttt{TensorIKKBZ}: Run of the algorithm for the precedence graph of $T^{[4]}$. Leg sizes are specified in the original tree tensor network, while the scores are shown as \emph{symbolic} fractions in the upcoming steps, e.g., $\sigma(T^{[3]}) = \cf{5}{4}$. Contradictory sequences are highlighted when a normalization is due. In the last step, we obtain the optimal linear contraction order for the precedence graph of $T^{[4]}$.}
    \label{fig:ikkbz_exec}
\end{figure}

An execution of the algorithm is visualized in Fig.~\ref{fig:ikkbz_exec}. We are considering only the precedence graph of $T^{[4]}$, but the algorithm would have to consider all precedence graphs. In the first phase, the algorithm builds the precedence graph of $T^{[4]}$, which is equivalent to rooting the tree tensor network in $T^{[4]}$. The subtree of $T^{[2]}$ is not yet linearized, thus the algorithm merges the linearizations of $T^{[1]}$ and $T^{[5]}$, which in this case are the nodes themselves, by observing the score monotonicity w.r.t. ``$\leq$'' (as defined in Thm.~\ref{theorem:asi_cost_function}). After the merge step, a normalization step is required (step~\fbox{\scriptsize 3}), since $T^{[2]} \rightarrow T^{[1]}$, but $\sigma(T^{[1]}) \leq \sigma(T^{[2]})$. This creates the compound tensor $T^{[2]}T^{[1]}$ which comprises both $T^{[2]}$ and $T^{[1]}$, in that order. Its score is calculated as described in Alg.~\ref{algo:normalization}.

Afterward, we are able to linearize $T^{[4]}$ and, in the final step, denormalize it, i.e., resolve the compound tensors into their constituent tensors. In our case, the compound tensor $T^{[2]}T^{[1]}$ is decomposed into individual tensors, as in step \fbox{\scriptsize 5}. Finally, we obtain the optimal linear contraction order for the precedence graph of $T^{[4]}$. That is, $T^{[4]}$ should be contracted first with $T^{[3]}$, then with $T^{[2]}$, followed by $T^{[1]}$ and $T^{[5]}$.

\section{Experiments}

\subsection{Graphical Notations}\label{appendix:graphical}

\begin{figure}
    \centering
    \begin{subfigure}[b]{.5\textwidth}
        \centering
        \includegraphics[]{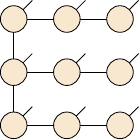}
        \caption{Fork-Tensor Product State~\cite{ftps}}
    \end{subfigure}%
    \begin{subfigure}[b]{.5\textwidth}
        \centering
        \includegraphics[]{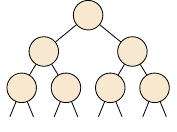}
        \caption{Hierarchical Tucker}
    \end{subfigure}
    \caption{Graphical notations}
    \label{fig:appendix_graphical_new}
\end{figure}

In Sec.~\ref{sec:evaluation}, we benchmarked on the Hierarchical Tucker and the Fork-Tensor Product State (FTPS)~\cite{ftps}, both tree tensor networks. Their graphical notations are both illustrated in Fig.~\ref{fig:appendix_graphical_new}.

\subsection{Sycamore}\label{appendix:sycamore}

\begin{figure}
    \centering
    \includegraphics[width=0.9\textwidth]{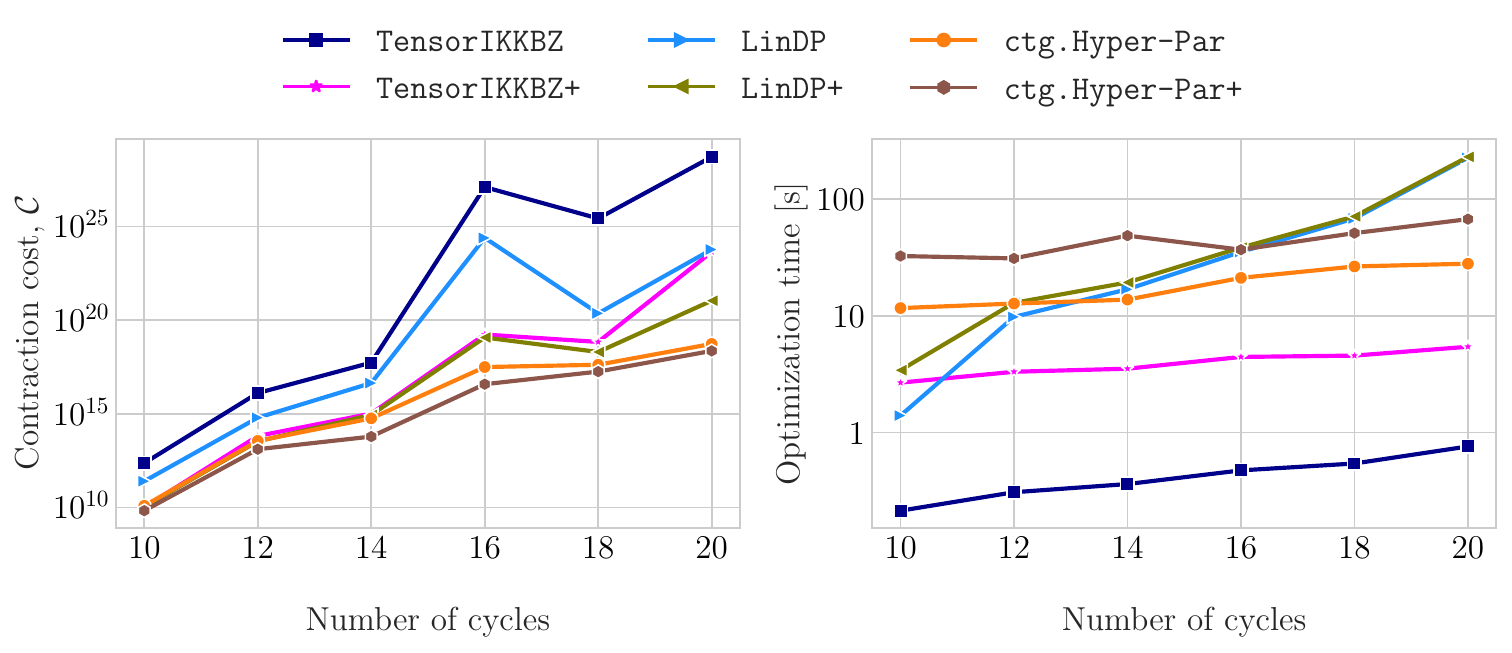}
    \caption{Benchmark: Sycamore circuit~\cite{sycamore}. We suffix with ``+'' the optimizers for which we enabled local optimization~\cite{huang2020classical} in \texttt{cotengra}.}
    \label{fig:sycamore_exp}
\end{figure}

The Sycamore architecture is the one used and executed in the quantum supremacy experiment of Arute et al.~\cite{sycamore}. In our experiments, we enable local optimization~\cite{huang2020classical} and suffix the optimizers where we have enabled it with ``+''. For \texttt{ctg.Hyper-Par}, we use the default value for the number of samples, 128. Note that in this setting, the underlying tensor network is not tree-shaped. Hence, as discussed in Sec.~\ref{subsec:gtn}, we need to choose a spanning tree of the tensor network. In Fig.~\ref{fig:sycamore_exp}, we plot both the contraction costs and the optimization times.

First, since the underlying tensor network of the Sycamore circuit is a generic one, our algorithms cannot provide any optimality guarantee (but rather only on the minimum spanning tree provided). Second, we observe that for a small number of cycles, namely less than or equal to 14, \texttt{LinDP} is on par with \texttt{ctg.Hyper-Par}. However, as the number of cycles increases, the contraction costs offered by \texttt{LinDP} become less competitive. Moreover, \texttt{LinDP} suffers from its quartic time complexity. Since our main technical result is the \texttt{TensorIKKBZ} algorithm, we leave it as future work to further improve both the contraction costs and optimization times of \texttt{LinDP} in the context of tensor networks.

\end{document}